\documentclass[a4paper,10pt,twocolumn]{article}
\usepackage[latin1]{inputenc}
\usepackage[english]{babel}
\usepackage[dvips]{graphicx}
\usepackage{amsmath,amsfonts}
\usepackage{amssymb}
\usepackage{fancyhdr}
\usepackage{afterpage}
\usepackage{longtable}
\usepackage{mathrsfs}
\usepackage{sidecap}

\newcommand\hspacetable{\rule{0pt}{2.6ex}}
\newcommand\hspaceunder{\rule[-1.2ex]{0pt}{0pt}}

\begin{document}

\title{State-of-the-art models for the phase diagram of carbon and diamond nucleation}

\author{L. M. Ghiringhelli$^{1}$\footnote{Current address:
Max-Planck-Institute for Polymer Research, Ackermannweg 10, 55128
Mainz, Germany}~, C. Valeriani$^{2}$\footnote{Current address:
School of Physics, James Clerk Maxwell Building, King's Buildings,
University of Edinburgh, Mayfield Road, EH9 3JZ, Edinburgh, UK}~,
J. H. Los$^3$, E. J. Meijer$^1$, \\ A. Fasolino$^{1,3}$, and D.
Frenkel$^{1,2}$\footnote{Current address: Dept. of Chemistry,
University of Cambridge, Lensfield Road, Cambridge CB2 1EW, UK}\\
\vspace*{0.1cm}\\
{\small $^1$ van 't Hoff Institute for Molecular
Sciences,Universiteit van Amsterdam,} \\ {\small Nieuwe
Achtergracht 166, 1018 WV Amsterdam, The
Netherlands.}\\
{\small $^2$ FOM Institute for Atomic and Molecular Physics,}\\
{\small Kruislaan 407, 1098 SJ Amsterdam, The Netherlands.} \\
{\small $^3$ Institute for Molecules and Materials, Radboud
University Nijmegen}\\ {\small Heyendaalseweg 135, 6525 AJ
Nijmegen, The Netherlands}}

\maketitle

\begin{abstract}
We review recent developments in the modelling of the phase
diagram and the kinetics of crystallization of carbon. In
particular, we show that a particular class of bond-order
potentials (the so-called LCBOP models) account well for many of
the known structural and thermodynamic properties of carbon at
high pressures and temperatures. We discuss the LCBOP models in
some detail. In addition, we briefly review the ``history'' of
experimental and theoretical studies of the phase behaviour of
carbon. Using a well-tested version of the LCBOP model (viz.
LCBOPI$^+$) we address some of the more controversial hypotheses
concerning the phase behaviour of carbon, in particular: the
suggestion that liquid carbon can exist in two phases separated by
a first-order phase transition and the conjecture that diamonds
could have formed by homogeneous nucleation in Uranus and Neptune.
\end{abstract}

\section{Introduction}
Carbon exhibits a rich variety of solid structures: Some are
thermodynamically stable, most are not. To be specific: solid
carbon can be found in the two well-known crystalline phases,
diamond and graphite, and in amorphous states, such as glassy
carbon and carbon black. Furthermore, the existence of additional
metastable solid phases at relatively low pressure, the so-called
carbynes, is still hotly debated~\cite{Whittaker,SmithBuseck}). In
addition to the bulk phases, there are the more recently
discovered fullerenes, C$_{60}$ and C$_{70}$~\cite{Kroto},
nanotubes~\cite{Iijima}, and graphene \cite{Novoselov}.\\
The reason why a simple element such as carbon can manifest itself
in so many different forms is related to its unusual chemical
properties: carbon exhibits three different possibilities for
covalent bond formation: $sp_3$ hybridization appears in diamond,
$sp_2$ hybridization is found in graphite, graphene, nanotubes,
and fullerenes, whilst in carbynes, C should
exhibit $sp$ hybridization.\\
Because of their high cohesive energies and concomitant high
activation energies that must be overcome in  structural phase
transformations, carbon polymorphs often exist in metastable form
well inside pressure-temperature regions where another solid form
is thermodynamically stable. For example, it is well known that
diamonds survive at normal P-T conditions, where graphite is the
thermodynamically stable phase. Conversely, graphite tends to
persist at very high pressures, deep into the diamond stability
region of the phase diagram.\\
It is also interesting that, at zero pressure and temperature,
graphite and diamond have a very similar (and quite large) binding
energy \textit{per atom}, i.e. 7.37 eV (graphite) vs 7.35 eV
(diamond). This fact might suggest (and it has indeed
been suggested) that also in disordered phases like the liquid, the two
local structures -- graphite-like and diamond-like -- could
compete. In fact, as we discuss below, the possibility of the
existence of two distinct and partially immiscible liquid phases
of carbon has been a subject of much debate.

In a liquid--liquid phase transition (LLPT), a liquid substance
displays an abrupt change in some local  or global  property
within a narrow band of pressures and temperatures. Local
properties that may change in a LLPT are the local coordination or
hybridization, typical global properties  that are affected are
the density or the resistivity. LLPT's in dense, atomic liquids
are typically difficult to probe experimentally: the candidate
transitions often occur at extreme pressures and/or temperatures
or appear in metastable regions of the phase diagram (and may be
hidden by competing solidification). Evidence for LLPT's have been
found for a number of atomic systems, such as Cs~\cite{Jayaraman},
As~\cite{Bellisent}, Bi~\cite{YoungBook}, Ge~\cite{GeLLPT},
Hg~\cite{HgLLPT}, S~\cite{SLLPT}, Sb~\cite{SbLLPT},
Se~\cite{SeLLPT}, Si~\cite{Angell,SiLLPT}, Sn~\cite{SnLLPT},
H$_2$~\cite{H2}, I$_2$~\cite{I2},
N$_2$\cite{RossRogers2006,Mukherjee2007}. The best established
experimental example of a LLPT in an atomic liquid is the case of
phosphorus.  A transition between a fluid of tetrahedral $P_4$
molecules and a network forming (and metallic) liquid was
predicted on theoretical grounds~\cite{Pauling,HohlJonesMD}, and
subsequently verified experimentally~\cite{KatayamaN,Monaco}. The
LLPT in phosphorus has been analysed in several numerical
studies~\cite{Morishita,Senda,GhiringhelliP,Ghiringhelli05PT}.
Many other network-forming liquids are also expected to exhibit
LLPT'a: first and foremost water \cite{Mishima98,StanleyN}, but
also SiO$_2$~\cite{PoolePRL97} and GeO$_2$~\cite{Itie}. Although
considerable progress has been made in the theoretical description
of LLPT's~\cite{Rapoport,
Kittel,Aptekar,Korsunskaya,Brazhkin,Tanaka00}, a unified
theoretical picture is still lacking.

In this review we discuss the phase diagram of solid and liquid
carbon at high pressures and temperatures on the basis of the
results of numerical simulations; both quantum and classical. We
present evidence that the presence of graphite-like and
diamond-like local structures in the liquid does not give rise to
liquid-liquid demixing but that the predominant local structure in
the liquid varies strongly with pressure. The fact that the liquid
is locally either graphite-like or diamond-like has dramatic
consequences for the nucleation of the diamond phase, a finding
that may have some consequences for our understanding of
carbon-rich planets or stars.  Wherever possible, we discuss our
own results in the context of  the relevant literature about the
carbon phase diagram, about a possible LLPT in this system and
about the possibility of diamond formation in planetary interiors.

In section~\ref{LosFasolino} we give an overview of the bond-order
potential (``LCBOP'') that was used to compute both the
equilibrium phase diagram of carbon and the pathway for diamond
nucleation in liquid carbon.  We also discuss in some detail the
different variants of the  LCBOP potential
\cite{LosFasolino,LCBOPII-I} and explain the rationale behind the
choice of the present LCBOP potential.

In section \ref{history} we briefly summarize some of the earlier
ideas about the phase diagram of carbon (in particular, about the
crystalline phases and the liquid). We pay special attention to
the slope of the diamond melting curve and the (possible)
heating-rate dependence of the graphite melting curve.

In section \ref{ejm} we report our results concerning the phase
diagram, in  the context of recent  first-principle simulations
and experiments.

In section \ref{LLPT} we review the arguments that have been put
forward to support the idea that liquid carbon can undergo a LLPT.
We argue that, to the extent that we can trust the present models
of liquid carbon, a LLPT in carbon is not be expected.

In section \ref{dianucl} we discuss our numerical results
concerning the (homogeneous) nucleation of diamond from the bulk
liquid. In particular, we discuss in some detail the numerical
approach that was used in Ref.~\cite{Ghiringhelli05nucl}. In
addition,  we focus on the structural analysis of the small solid
clusters in the liquid and we discuss the implications of our
findings for the formation of diamonds in carbon-rich star systems
and the interior of giant planets.

\section{The LCBOP-family}
\label{LosFasolino} Whilst the crystalline phases of carbon can be
simulated by traditional force fields that do not allow
coordination changes, a study of the liquid phase and, a fortiori,
of  phase transformations between phases with different local
coordination, requires a potential that can describe carbon in
different coordination states. The LCBOP potential was designed
with this objective in mind. LCBOP stands for "Long range Carbon
Bond Order Potential", and represents a bond-order potential for
pure carbon that includes long-range (LR) dispersive and repulsive
interactions \cite{LosFasolino} from the outset. We stress that
LCBOP is not based on an existing short-range (SR) bond-order
potential to which LR interactions have been added a posteriori,
although such an approach has been proposed in the
literature~\cite{Sinnott,Petukhov,Stuart}. The latter approach
requires a rather special procedure to avoid interference with the
SR potential and suffers from a loss of accuracy. In the LCBOP
these problems are circumvented in a natural way.

We note that the term "long range" may be confusing in this case,
as the cut-off of the LR potential in LCBOP is only 6 \AA.
Usually, the term "long range" is only used for interactions with
a much longer range, such as Coulomb interactions. Here we use it
as a synonym for "non--bonded", referring to a range much larger
than the typical distances between chemically bonded atoms.

After the introduction of LCBOP in Ref. \cite{LosFasolino}, a
number of significant modifications have been introduced in order
to improve its description of all carbon phases, including liquid
carbon. To facilitate the distinction between the different LCBOP
potentials, the various versions have been named LCBOPI,
LCBOPI$^+$ and LCBOPII. LCBOPI, introduced as LCBOP in Ref.
\cite{LosFasolino}, does not include torsion interactions. As
torsion interactions were shown to play an important role in
liquid carbon \cite{Wu2002}, we introduced a refinement of LCBOPI,
called LCBOPI$^+$, when we performed our first study of liquid
carbon~\cite{Ghiringhelli04}. LCBOPI$^+$, includes, among other
changes, conjugation dependent torsional interactions. Clearly,
describing the liquid phase requires a robust form of the
potential in order to deal with configurations that are quite
unlike the regular topologies in crystal lattices. LCBOPII
addresses this problem: it includes several important improvements
over LCBOPI$^+$. An important innovation in LCBOPII is the
addition of so-called middle range (MR) interactions, introduced
to bridge the gap between the extent of the tail of the covalent
interactions as found in ab-initio calculations, (up to 4.5 \AA$~$
in certain cases) and the rather short cut-off of only 2.2 \AA$~$
in LCBOPI$^+$ for these interactions.

In the remainder of this section we give a brief step-by-step
description of the transition from bond-order potentials (BOPs) to
LCBOPI$^+$. All the results presented in later sections,
concerning the phase diagram, the liquid structure, and the
nucleation issues are obtained with this version of the potential.
In Appendix A we discuss the LCBOPII potential, with a short
account of results obtained with this refined version of the
potential. We aim at giving the flavour of the potentials in a
mainly descriptive way with graphical illustrations, minimizing
mathematical formulation.

\subsection{Bond-order potentials}
A bond-order potential (BOP) is a reactive potential, i.e. able to
deal with variable coordination. It provides a quantitative
description of the simple idea that the bonds of an atom with many
neighbours are weaker than those of an atom with few neighbours,
as the cohesive ability of the available electrons has to be
shared among the neighbours. For carbon, each atom delivers four
valence electrons. If these four electrons have to make the six
bonds in a simple cubic lattice, then it is evident that each of
these bonds is  weaker than a bond in the diamond or graphite
lattice with coordinations 4 and 3 respectively. On the other
hand, the number of bonds is larger for the simple cubic lattice.
So there is a balance to be made, which in the case of carbon has
the result that  graphite is the most stable phase at ambient
pressure.

A thorough analysis of these bonding properties, based on a
quantum mechanical description, has been given by Anderson
\cite{Anderson,Anderson2,Anderson3} and Abell \cite{Abell}. In
their description, that forms the basis of the tight binding
models, the electronic wave function is approximated as a sum of
localized atomic orbitals. Abell showed that for a regular
lattice, i.e. with an identical environment for each atom, and
within the assumption that the overlap integral for orbitals on
different atoms is non-vanishing only for nearest neighbours, the
binding energy per atom is given by:
\begin{equation}
E_b =  \frac{1}{2} Z ( q V_\mathrm{R} (r) + b V_\mathrm{A} (r) )
\label{eb1}
\end{equation}
where $ Z $ is the number of nearest neighbours, $q$ is the number
of valence electrons per atom, $ V_\mathrm{R} (r)$ is a two-body
potential describing the core repulsion, $ V_\mathrm{A} (r)$ is a
two-body attractive potential, and $b$ is the so-called bond
order, a many-body term dependent on the local environment of the
atoms. Abell also showed that the coordination dependence of $b$
is fairly well approximated by:
\begin{equation}
b = b(q,Z) = \alpha (q) Z^{-1/2} \label{b1}
\end{equation}
with $\alpha (q)$ a function of $q$, specified in Ref.
\cite{Abell}. Assuming exponential functions $ V_\mathrm{R} (r)= A
\exp(- \theta r) $ and $ V_\mathrm{A} (r)= - B \exp(- \lambda r)
$, with $A$, $B$, $\lambda$, and $\theta$ fitting parameters, as a
reasonable approximation for overlapping atomic orbitals from
atoms at distance $r$, and defining $ S = \theta/\lambda $, some
algebra leads to a total binding energy given by:
\begin{eqnarray}
\nonumber E_b &=& B \alpha (q) \ \frac{S-1}{2S} \left( \frac{ B
\alpha(q)}{ qAS } \right)^{\frac{1}{S-1}} Z^{\frac{S-2}{2(S-1)}} \\
&=& C Z^{\frac{S-2}{2(S-1)}} \label{eb2}
\end{eqnarray}
and an equilibrium nearest neighbour distance given by:
\begin{equation}
r_{eq} = \frac{1}{\theta - \lambda} ln \left( \frac{qAS
\sqrt{Z}}{B \alpha(q) } \right) = \frac{1}{2( \theta - \lambda)}
ln Z + C' \label{req2}
\end{equation}
where $C$ and $ C'$ are constants. Eq. \ref{eb2} implies that for
$ S > 2 $ high coordination structures (close packing) are
favoured (metals), whereas for $S < 2$ the dimer will be the most
stable structure (extreme case of covalent bonding). As, in
general, the repulsion falls off (much) faster than the
attraction, i.e. $ \theta > \lambda ~(S>1)$, Eq. \ref{req2}
implies that $ r_{eq} $ is monotonically  increasing with
coordination. Combining Eqs. \ref{eb2}  and \ref{req2} yields a
simple relation between $ E_b $ and $ r_{eq} $, namely $ E_b
\propto \exp{(\theta - 2 \lambda) r_{eq }}$.

In principle BOPs are based on the above bonding ideas. However,
the transferability to different types of structures and materials
has been greatly enhanced by a quite reasonable extension in the
functional form of the bond order $b$. The simplest bond order
$b_{ij}$ for a bond $ij$ according to a BOP in the style of
Tersoff~\cite{Tersoff4} and Brenner~\cite{Brenner90} reads:
\begin{equation}
b_{ij} = \alpha \left( 1 + \sum_{k \neq i,j} G ( \theta_{ijk} )
\right)^{\epsilon} \label{b3}
\end{equation}
where the sum runs over the nearest neighbours other than $j$ of
atom $i$, $ G ( \theta_{ijk} ) $ is an adjustable function of the
bond angles $ \theta_{ijk} $ and $ \epsilon $ is a negative
exponent but not necessarily -1/2. Taking a constant $ G (
\theta_{ijk} ) = 1 $ and $ \epsilon = -1/2 $ yields $ b = \alpha
Z^{-1/2} $, i.e. one recovers the functional form Abell found for
$b$ (Eq. \ref{b1}). This form includes the effect of bond angles
in a natural way, and has the ability to fit a large set of data
quite well, explaining the success of BOPs.

\subsection{LCBOPI}
\begin{figure}[b!] \centering
\includegraphics[width=0.80\columnwidth,clip]{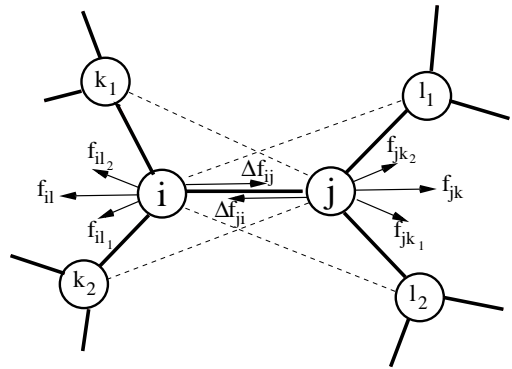}
\protect \caption{Schematic representation of the LCBOP approach
to the inclusion of LR interactions. To preserve the right
equilibrium bond distance for a given bond $ij$, the repulsion of
atom $i$ due to the added LR interactions with the atoms $l_1$ and
$ l_2 $ (represented by the force $ {\bf f}_{il} = {\bf f}_{il_1}
+ {\bf f}_{il_2} $)  is compensated by a stronger attractive part
in the SR interaction between $i$ and $j$, represented by the
extra force $ \Delta f_{ij} $. The same holds for atom $j$. For
convenience, here only the main, repulsive LR interactions are
drawn, whereas in reality this compensation is for the sum of all
LR interactions.} \label{figrev1}
\end{figure}

The main innovative feature of LCBOPI \cite{LosFasolino} concerns
the treatment of the LR van der Waals interactions. One of the
challenges here is to add LR interactions which describe not only
the interlayer graphitic binding but also the rather strong $
\pi$-bond repulsion in graphite for decreasing interlayer distance
without paying a price in the accuracy of the covalent binding
properties. A correct description of these interactions requires a
LR potential, $ V^{lr} (r) $ that is repulsive in the distance
range corresponding to the second nearest neighbours (in diamond
and graphite). In the LCBOP-family, to get the right equilibrium
lattice parameter, this extra repulsion has been compensated by a
somewhat stronger attractive part of the covalent interaction,
achieved by an appropriate parametrisation of the SR potential.
This is schematically illustrated in Fig. \ref{figrev1}.

\begin{figure}[t!]
\centering
\includegraphics[width=0.85\columnwidth,clip]{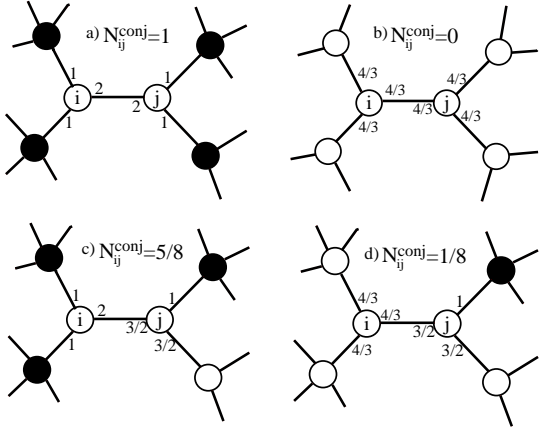}
\protect \caption{Schematic presentation of the interpolation
approach in the determination of $ N^{conj}_{ij} $ for mixed
coordination situations (c and d). Black atoms are saturated
atoms, white atoms are unsaturated. The numbers near the atoms $i$
and $j$ indicate the distribution of electrons among their bonds.
Each single bond, i.e. a bond with a saturated neighbour, takes 1
electron. The remainder of the electrons (in total 4) are equally
shared among the bonds with unsaturated atoms. A linear dependence
of $ N^{conj}_{ij} $ on the total number of electrons involved in
the bond $ij$ with $ 0 \leq N^{conj}_{ij} \leq 1 $ as an
additional constraint leads to the given values for $
N^{conj}_{ij} $.} \label{figrev2}
\end{figure}

Another feature of LCBOPI is that it contains a reasonable,
physically motivated interpolation scheme for the conjugation term
to account for a mixed saturated and unsaturated environment. In
this approach each atom supplies a number of electrons to each of
its bonds with neighbouring atoms according to a certain
distribution rule, the total sum being equal to the valence value
4 of carbon. The character of a certain bond $ij$, and its
conjugation number $ N^{conj}_{ij} $, a number between 0 and 1
quantifying effects beyond nearest neighbours, is determined by
the sum of the electrons supplied by atom $i$ and atom $j$. The
interpolation model is further illustrated in Fig. \ref{figrev2}.

\subsection{LCBOPI$^+$}
The potential LCBOPI$^+$ is given by LCBOPI supplemented with
torsion interactions and a correction of the angle dependent part
of the bond order for configurations involving low coordinations
and small angles. Similar modifications were also included in the
REBO potential \cite{BrennerREBO}, although the torsion term
contains a significant difference. For LCBOPI$^+$, following the
results of ab-initio calculations, the shape of the torsion energy
curve as a function of the torsion angle depends on conjugation,
whereas for the REBO potential the curves for a double and a
graphitic bond are equally shaped but scaled (see top panel of
Fig. \ref{figrev3}). Details on LCBOPI$^+$ are given in the
appendix A of Ref.~\cite{LCBOPII-II} and in
Ref.~\cite{Ghiringhelli-thesis}.

\section{The phase diagram of carbon at very high pressures and
temperatures} \label{history} In this section we  give a review of
experimental and theoretical works aimed at determining the phase
behaviour of carbon at high temperatures and pressure. We  follow
a ``historical'' approach, starting from the beginning of the
twentieth century, up to the most recent results coming from
experiments and computer simulations. A historical approach may
give a better understanding  why certain issues have been, and in
some case still are, controversial. After setting the stage (with
a particular attention to the ideas about the sign of the slope of
the diamond melting curve and the long debated issue of the
position and nature of the graphite melting curve), we  focus on
the topic of the LLPT for carbon.

\subsection{The history of carbon phase diagram} \label{CPDhistory}
\begin{figure}[!t]
\begin{center}
\includegraphics[width=0.8\columnwidth,clip]{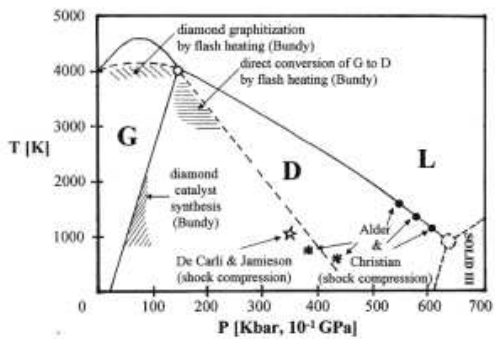}
\caption{The phase diagram of carbon at high pressures proposed by
Bundy in 1963 ~\cite{BundyGDCL}.} \label{BundyPD}
\end{center}
\end{figure}

\begin{figure}[!t]
\begin{center}
\includegraphics[width=0.8\columnwidth,clip]{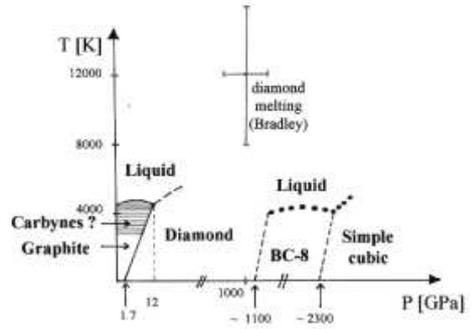}
\caption{In 2004, a schematic representation of the phase diagram
of carbon at high pressures would have looked more or less as
indicated in this figure. Full curves correspond to phase
boundaries for which thermodynamic data are available. More recent
developments are discussed in the next section.}
\label{up-to-nowPD}
\end{center}
\end{figure}

One of the earliest phase diagrams of carbon appeared at the
beginning of the twentieth century, and is due to H. Bakhuis
Roozeboom~\cite{Roozeboom}, who estimated the phase behavior of
carbon on the basis of thermodynamic arguments. Of the two solid
phases, diamond was recognized to have a slightly greater vapor
pressure at a given temperature. The temperature of the
graphite/liquid/vapor triple was believed to be around 3000~K. In
1909 Tamman~\cite{Tamman} postulated the existence of a region
where graphite and diamond are in pseudo-equilibrium. The
existence of this pseudo-equilibrium region was at the basis of
the method of synthesizing diamond starting from carbon saturated
solutions of molten iron, silver, or silicates. In 1938, Rossini
and Jessup~\cite{Rossini} of the U.S. Bureau of Standards used
accurate thermodynamic data to estimate that at 0~K the lowest
pressure at which diamond would be stable against graphite is
around 1.3~GPa, and around 2~GPa at 500~K. In 1939, the Russian
scientist Leipunskii~\cite{Leipunskii} published a review of the
problem of diamond synthesis. On the basis of thermodynamic data,
he suggested that the melting curve of graphite might be at about
4000~K, with possibly some increase with pressure. This value for
the melting  of graphite was rather well verified the same year by
Basset~\cite{Basset}, who established the graphite/liquid/vapor
triple point to be at about 11~MPa and 4000~K. In that same
publication, Basset reported on a rather pressure independent
melting temperature of graphite at $\sim 4000$~K, from atmospheric
pressure up to 0.1 GPa. In 1947 Bridgman~\cite{Bridgman} addressed
the problem of extrapolating the graphite/diamond coexistence
curve beyond the region where it can be estimated from known
physical properties (4~GPa/1200~K). He concluded that there was a
possibility that at higher temperatures the rate of increase of P
with T along the curve would decrease. This hypothesis was later
supported by Liljeblad~\cite{Liljeblad} in 1955, while Berman and
Simon~\cite{Berman} in the same year came to the conclusion that
the best extrapolation would be a straight line. Experiments that
could decide this issue were started by Bundy and coworkers in
1954, when they accomplished diamond synthesis by activating the
graphite-to-diamond reaction with the use of different
solvent-catalyst metals. The relevant experimental  data were
published only much later~\cite{BundyNat,BundyGML}, and are
compatible with the
Berman-Simon straight line extrapolation.\\
Bundy and his group made also extensive experiments on graphite
melting at pressures much higher than the graphite/liquid/vapor
triple point. The determination of the graphite melting curve is
an experimental challenge for several reasons. First of all, to
reach pressures as high as 10~GPa, the sample must be in direct
contact with a solid container and, because the melting
temperature are so high, this container must be made of a material
that is as refractory and inert as possible (Bundy chose boron
nitride, pyrophyllite, MgO and diamond powder). In addition, both
the heating of the sample and the observations of the
high-pressure/high-temperature phase must be carried out very
rapidly, before the wall material can melt or react with the
carbon sample. The experiments were performed by discharging an
electrical capacitor through the sample (this procedure is known
as flash heating), and by monitoring the current through, and the
voltage across it by means of a two-beam oscilloscope. The
discharge circuit was designed to have energy insertion in the
sample within a few milliseconds. The interpretation of such
experiments, is rather sensitive to the assumed pressure and
temperature dependence of the material under study and on the
assumption that the pressure of the graphite specimen during rapid
heating is the same as in  a quasi-static process. With these
assumptions, Bundy' s experiments gave a graphite melting curve as
shown in Fig.~\ref{BundyPD}. A maximum melting temperature of
about 4600~K was detected in the region of 6~GPa to 7~GPa. The
presence of a region with a negative d$T/$d$P$ along the melting
curve indicates that, at those pressures, the density of the
liquid at the melting temperature is greater than that of the
solid.\\
Interrupting for a moment the historical order of events, we note
that, throughout the past century, different experiments located
the graphite melting curve at rather different temperatures
~\cite{Pirani,Basset,Schoessow,Heremans,BundyGML,WhittakerG,ShanerG,MillerGML,Pottlacher,BundyC,TogayaPRL,Musella}.
At low pressure, the melting temperature (T$_m$) was found at
values ranging from $\sim 3800$ to $\sim 5000$ K. Asinovskii {\em
et al.}~\cite{Asinovskiy97} pointed out the non negligible
dependence of graphite $T_m$ on the heating rate of the sample.
Specifically, heating times of the order of 10$^{-5}$~s
~\cite{ShanerG,Pottlacher} yielded estimates of $T_m \sim
4800-5000$~K; heating times of the order of
10$^{-3}$~s~\cite{MillerGML,TogayaPRL} suggested $T_m \sim
4500-4600$~K; finally experiments with heating times of the order
of one second~\cite{Basset,WhittakerG} were consistent with the
assumption that $T_m \sim 3800-4000$~K. In
ref~\cite{Asinovskiy97}, after a thorough discussion of the
experimental methods, the authors recommended that only data
coming from experiments with heating time of the order of seconds
or more should be accepted. This implied that most of the
available data on graphite melting had to be reconsidered and that
the question of the position and the shape of the melting curve is
still open. On the basis of a series of laser induced slow heating
experiments~\cite{Asinovskiy97} (i.e. heating times of the order
of one second), Asinovskii {\em et al.} proposed the triple point
vapor/liquid/graphite at $\sim 4000$~K and 0.1~MPa (i.e.
atmospheric pressure), in clear contradiction to the commonly
accepted values~\cite{BundyC} of $\sim 5000$~K and 10~MPa. The
next year the same authors~\cite{Asinovskiy98} published results
concerning the position of the graphite melting curve. With ohmic
heating of graphite samples at heating rates of about
100~K/minute, they found $T_m = 3700$~K at 0.25~MPa (typically,
samples melted after
one hour of steady heating).\\
Coming back to Bundy's work, during the experiments on graphite
melting Bundy and his group also investigated the graphitization
of diamond by flash-heating under pressure. Small diamond crystals
where embedded in the graphite sample, pressurized and then
flash-heated. Experiments indicated that there is a sharp
temperature threshold at which the diamond crystals completely
graphitized. This threshold is a few hundreds degrees lower than
the graphite melting curve.\\
Attempts to obtain direct (i.e. without resorting to a catalyst
material) conversion of graphite into diamond by the application
of high pressure date back to the beginning of the twentieth
century. Success came only in 1961, when De Carli and
Jamieson~\cite{DeCarli} reported the formation and retrieval of
very small black diamonds when samples of low-density
polycrystalline graphite were shock compressed to pressures of
about 30~GPa. Later in 1961 Alder and Christian~\cite{Alder}
reported results on the shock compression of graphite that were in
substantial agreement with those of De Carli and Jamieson.\\
Bundy~\cite{BundyGDCL} achieved direct conversion of graphite into
diamond by flash-heating graphite sample in a static pressure
apparatus, at pressures above the graphite/diamond/liquid triple
point. The threshold temperature of the transformation was found
several hundred degrees below the melting temperature of the
graphite, and decreasing at higher pressures. The phase transition
was revealed by a sharp drop in the electrical conductivity of the
samples (that were retrieved as pieces of finely
polycrystalline black diamond).\\
By linking his own results with earlier experimental and
theoretical findings, Bundy~\cite{BundyGDCL} proposed in February
1963 a phase diagram of carbon at high pressures that is
illustrated in Fig.~\ref{BundyPD}. The diamond melting curve was
believed to have negative slope by analogy with the other Group IV
elements~\cite{Jamieson,Bundy69}), and on the basis of evidence
collected during the experiments of Alder
and Christian~\cite{Alder}.\\
In 1973 Van Vechten ~\cite{VanVechten} predicted the phase diagram
of carbon by rescaling the behavior of other Group IV elements
that are experimentally more accessible, using the
electronegativity as a scale parameter. In 1979 Grover
~\cite{Grover} calculated a phase diagram by using a
phenomenological equation of state for the description of various
solid and liquid phases of carbon. He used physically motivated
approximations for the free energies of the various phases, with
parameters adjusted to match the available data on the equations
of state. He concluded that, at all pressures, diamond transforms,
before
melting, into a solid metallic phase.\\
On the basis of experimental evidence~\cite{Goresy}, in 1978
Whittaker~\cite{Whittaker} proposed the existence of a novel
crystalline phase for elemental carbon, called ``carbyne''. The
stability region of carbyne is sketched in Fig.~\ref{up-to-nowPD})
and the structure of this phase (though expected in different
allotropes) is generally that of a chains of alternated single and
triple bonds, i.e. $(-$C$\equiv$C$-)_n$, arranged in a hexagonal
array bundled by dispersion interactions. The existence of a
carbyne form was later questioned by Smith and
Buseck~\cite{SmithBuseck}, who claimed that all the experimental
evidence could also be accounted for by the presence of sheet
silicates. This dispute continued to this day: the experimental
evidence for the
existence of carbyne is still being debated.\\
In recent years, the experimental effort has focused on the
collection of reliable data at even higher pressures, and on the
investigation of the properties of the different phases of carbon
at high temperatures and pressures. This challenging task has been
faced both with experiments and theory. On the experimental side,
the development of the diamond-anvil cell ~\cite{Jayaraman} for
high pressure physics has made it crucial to know the range of
stability of diamond under extreme conditions. The availability of
high-energy pulsed laser sources led to new tools for heating up
samples at very high temperatures (above the graphite melting
curve) ~\cite{Venkatesan}. These techniques were immediately
applied to the determination of the properties of liquid carbon
(i.e. whether it is a conducting metallic liquid or an insulator).
Unfortunately, due to the difficulties in interpreting the results
of these experiments, the nature of the
liquid state of carbon is still not characterized experimentally. \\
On the theoretical side, the appearance of ever more powerful
computers  made it possible to use electronic density-functional
(DF) theory~\cite{HK,KS} to predict the properties of materials
under extreme conditions. In 1983 Yin and Cohen ~\cite{Yin1}
studied the total energy versus volume and the free energies
versus pressure for the six possible lattices of carbon (fcc, bcc,
hcp, simple cubic, $\beta$-tin, diamond). The study was carried
out by using ab initio pseudopotential theory (this permits the
investigation of the properties of the atomic system at 0~K). Yin
and Cohen found that the calculated zero-pressure volume for
diamond is either close to or even smaller than those of the other
five phases. This is different from what is observed for the other
group IV elements, Si and Ge, and defies the common notion that
diamond is an open structure and should have higher specific
volume than the close packed solid structures. The relatively
dense packing of diamond would inhibit the phase transformations
at high hydrostatic pressures that are observed for heavier group
IV elements. In addition, it suggested a revision of the other
common notion that the diamond melting curve should have negative
slope, something that is to be expected when a liquid is denser
than the coexisting solid. Yin and Cohen also found that, at a
pressure around 2300~GPa, diamond converts to a simple cubic (sc)
phase. This work was later extended
~\cite{YinBC-8,BiswasBC-8,FahyBC-8} to consider also complex
tetrahedral structures. It was found that a distorted diamond
structure called BC-8 was stable versus diamonds at pressures
above 1000~GPa (see
Fig.~\ref{up-to-nowPD}).\\
In 1984 Shaner and coworkers ~\cite{Shaner} shock compressed
graphite and measured the sound velocity in the material at shock
pressures ranging from 80 to 140~GPa, and corresponding shock
temperatures ranging from 1500 to 5500~K.  They measured
velocities close to those of an elastic longitudinal wave in solid
diamond. These velocities  are much higher than those of a bulk
wave in a carbon melt. Since no melt was detected at pressures and
temperatures well above the graphite/diamond/liquid triple point,
the diamond melting curve should, according to these results, have
a positive slope. In 1990 Togaya ~\cite{TogayaDML} reported
experiments in which specimens of boron-doped semiconducting
diamond were melted by flash-heating at pressures between 6 and
18~GPa: these experiments provided clear
indications that the melting temperature of diamond increases with pressure.\\
In the same year ab-initio molecular dynamics (MD) studies
~\cite{GalliMartin-science} clearly showed that, upon melting
diamond at constant density, the pressure of the system increases.
These results imply that, at the densities studied, the slope of
the melting curve of diamond is positive. The shape of the diamond
melting curve has interesting consequences for the theory of
planetary interiors. Given our present knowledge of the phase
diagram of carbon and the existing estimates for the temperatures
and pressures in the interior of the outer planets Neptune and
Uranus, as well as in the Earth mantle, one might conclude that in
a large fraction of these planetary interiors the conditions are
such that diamond should be the stable phase of
carbon~\cite{Ross}; diamonds could then be expected to occur
wherever the carbon concentration is sufficiently high. In section
\ref{dianucl} we show that, when only homogeneous nucleation
is considered, our modelling predict that the driving force for
diamond nucleation is missing in giant planet interiors. In 1996
Grumbach and Martin \cite{CHiP} made a systematic investigation of
the solid and liquid phases of carbon over a wide range of
pressures and temperatures by using ab initio MD. These authors
studied the melting of the simple cubic and BC-8 solid phases, and
investigated structural changes in the liquid in the range
400-1000~GPa. They observed that the coordination of the liquid
changes continuously from about
four-fold to about six-fold over this pressure range.\\
In 2004, Bradley {\em et al.}~\cite{Bradley} reported experiments
on laser-induced shock compression of diamond up to 3000 GPa.
Through optical reflectivity measurements, they found for the
first time direct evidences of diamond melting, at an estimated
pressure of
$P = 1000 \pm 200$~GPa, and temperature $T = 12000 \pm 4000$~K.\\
Fig.~\ref{up-to-nowPD} summarizes the information about phase
diagram of carbon at the time when the present research was
started.

\subsection{Carbon phase diagram according to LCBOP} \label{ejm}
\textbf{Methods} We performed  Monte Carlo simulations on the
LCBOPI$^+$ model of carbon~\cite{LosFasolino,Ghiringhelli04} to
estimate the properties of the liquid, graphite, and diamond
phases of carbon. Coexistence curves were determined by locating
points in the $P-T$ diagram with equal chemical potential for the
two phase involved. For this purpose, we first determined the
chemical potential for the liquid, graphite, and diamond at an
initial state point ($P=$~10~GPa, $T=$~4000~K). This state point
is near the estimated triple point\cite{vanThielReePRB}.
Subsequently, the liquid-graphite, liquid-diamond, and
graphite-diamond coexistence pressures at $T=$~4000~K were
located. In turn, these coexistence points served as the starting
point for the determination of the graphite melting line, the
diamond melting line, and the graphite-diamond coexistence curve,
obtained by integrating the Clausius-Clapeyron equation (a
procedure also known as Gibbs-Duhem integration) \cite{IntegrCC}:
$\frac{\textrm{d}T}{\textrm{d}P}=\frac{T\Delta v}{ \Delta h}$
where $\Delta v$ is the difference in specific volume, and $\Delta
h$ the difference in molar enthalpy between the two phases.

We proceeded by first determining the Helmholtz free energy at a given
density and temperature by thermodynamic integration and subsequently
calculating the chemical potential using the procedure described in
Ref.~\cite{AnwarFrenkel}. Coexistence at a given $T$ is found at the
$P$ where the different $\mu$ cross.

For all phases, the Helmholtz free energy $F^\maltese$ of the
initial state point
 ($P=$~10~GPa,
$T=$~4000~K) was determined by transforming the system into a
reference system of known free energy  $F^\textrm{ref}$. The
transformation was imposed by changing the interaction potential:
$U_{\lambda}=(1-\lambda) U^\maltese + \lambda U^\textrm{ref}$.
Here, $U^\maltese$ and $U^\textrm{ref}$ denote the potential
energy function of the LCBOPI$^+$ and the reference system,
respectively. The transformation is controlled by varying the
parameter $\lambda$ continuously from $0$ to $1$. The free-energy
change upon the transformation was determined by thermodynamic
integration:
\begin{eqnarray}
F^\maltese &=& F^\textrm{ref} + \Delta F^{\textrm{ref}\rightarrow\,\maltese} =  \nonumber \\
&=& F^\textrm{ref} + \int_{\lambda=0}^{\lambda=1} d\lambda
\left\langle \frac{\partial U_{\lambda}}{\partial \lambda}
\right\rangle _{\lambda}= \nonumber \\
&=& F^\textrm{ref} + \int_{0}^{1} \textrm{d}\lambda \left\langle
U^\textrm{ref} - U^\maltese \right\rangle _{\lambda} \label{FE}
\end{eqnarray}
The symbol $\langle ... \rangle _{\lambda}$ denotes the  ensemble
average with the potential $U_{\lambda}$.

As reference system for the liquid we chose the well-characterized Lennard-Jones
(12-6) system, whilst the reference system for both diamond and
graphite was the Einstein crystal. General guidelines for these
kind of calculations are given in \cite{FrenkelSmit,AnwarFrenkel},
while a full description of the strategy adopted for the present
systems is given in \cite{Ghiringhelli-thesis}. The ensemble
averages needed for the thermodynamic integration were determined
from Monte Carlo (MC) simulations of a 216-particle system in a
periodically replicated simulation box. For simulations of the
graphite phase, the atoms were placed in a periodic rectangular
box with an initial edge-size ratio of about $1:1.5:1.7$. For the
liquid phase and diamond a periodic cubic box was used.

\textbf{From the Helmholtz free energy to the chemical potential}
The chemical potential $\mu$ along the $4000$~K isotherm was
obtained by integrating from the initial state point a fit,
$P(\rho)= a + b\rho + c \rho ^2$, through simulated $(P,T)$ state
points along the 4000~K isotherm.  Here, $\rho$ is the number
density, and $a$, $b$, and $c$ are fit parameters. This yields for
the chemical potential~\cite{AnwarFrenkel}:
\begin{equation}
\beta \mu (\rho) = \frac{\beta F^\maltese}{N} + \beta \left(
\frac{a}{\rho^\maltese} + b\
\textrm{ln}\frac{\rho}{\rho^\maltese}+b + c \left(2 \rho -
\rho^\maltese\right)
 \right)
\label{ES}
\end{equation}
Here, $\rho^\maltese$ denotes the number density at the initial
state point, $N$ the number of particles, and
$\beta=1/k_\textrm{B}T$, with $k_\textrm{B}$ the Boltzmann
constant.

\begin{figure}[t!]
\centering
\includegraphics[width=0.75\columnwidth,clip]{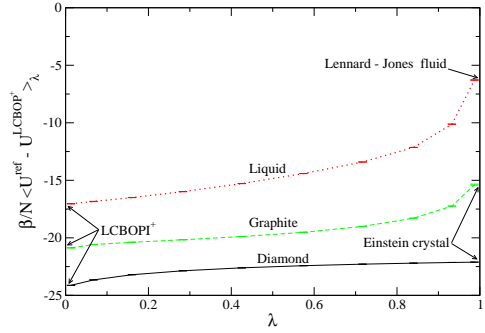}
\caption{Plots of the quantity $\beta / N\;\langle
U^{\textrm{ref}} - U^{\textrm{LCBOPI}^+} \rangle_{\lambda}$ (see
Eq.~\ref{FE}) as a function of the coupling parameter $\lambda$
for the liquid, graphite, and diamond. On the left side of the
horizontal axis ($\lambda=0$) is the pure LCBOPI$^+$, on the right
side ($\lambda=1$) is the reference system, i.e. the Lennard-Jones
liquid for the liquid phase and two Einstein crystals (with
different coupling constant) for graphite and diamond. The
simulated ten points per phase are marked by their error bars,
that are almost reduced to a single dash at this scale.}
\label{dUave}
\end{figure}

\textbf{Calculated coexistence curves} \label{s:resPD} The
equilibrium densities $\rho^\maltese/(10^3~\text{kg/m}^3)$ at the
initial state point ($P=$~10~GPa, $T=$~4000~K) were 3.425 for
diamond, 2.597 for graphite, and 2.421 for the liquid. Three
configurations at the equilibrium volume were then chosen as
starting state point for the free energy calculation.

\begin{table}[ht!]
\begin{center}
\begin{tabular*}{0.95\columnwidth} {@{\extracolsep{\fill}}l r r r}
\hline
&$a$ [GPa]\hspacetable\hspaceunder& $b$ [GPa~nm$^3$] & $c$ [GPa~nm$^6$]\\
\hline Liquid \hspacetable& 89.972 & $-1.9654$ & 0.011 092\\
Diamond &  74.809 & $-3.6307$&  0.019 102
\\
Graphite \hspaceunder& 108.29 & $-2.2707$ &  0.011 925 \\
\hline
\end{tabular*}
\end{center}
\caption{Parameters for the polynomial fitting of the 4000~K
isotherms of the three phases, according to: $P(\rho)= a + b\rho +
c \rho ^2$.} \label{EoS_fit}
\end{table}

\begin{figure}[!t]
\centering
\includegraphics[width=0.7\columnwidth,clip]{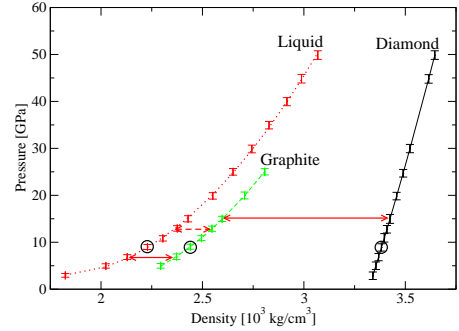}
\caption{Equations of state at 4000K for the liquid, graphite, and
diamond. The curves are quadratic polynomial fits to the data. The
circles indicate the points, at 10 GPa, for which the Helmholtz
free energy was determined using Eq.~\ref{FE}. The solid arrows
connect coexisting (stable) points, i.e. liquid/graphite and
graphite/diamond. The dashed arrow indicates the liquid/diamond
coexisting point, with graphite metastable.} \label{EoS_I+_4000K}
\end{figure}

The integrals related to the reference system transformation
(Eq.~\ref{FE}) were evaluated using a 10-point Gauss-Legendre
integration scheme. Fig.~\ref{dUave} shows the integrand $\langle
U^{\textrm{ref}} - U^{\textrm{LCBOPI}^+} \rangle_{\lambda}$ versus
$\lambda$. The smooth behaviour of the curves indicates that there
are no spurious phase transitions upon the transformation to the
reference system (the absence of such transitions is a necessary
condition for using this method). At the initial state point
($P=$~10~GPa, $T=$~4000~K), the calculated free energies($\beta
F^{\maltese}/N$) where $-24.824 \pm 0.006$, $-24.583 \pm 0.002$,
and $-25.137 \pm 0.002$, for graphite, diamond, and the liquid,
respectively.

Fig.~\ref{EoS_I+_4000K} shows the calculated state points along
the 4000~K isotherms for the three phases, along with the fitted
curves. The fit parameters are listed in Table \ref{EoS_fit}.
Employing subsequently Eq.~\ref{ES} we obtained the  calculated
chemical potential $\mu$ along the 4000~K isotherm for the liquid,
graphite, and diamond phase.
\begin{figure}[!t]
\centering
\includegraphics[width=0.75\columnwidth,clip]{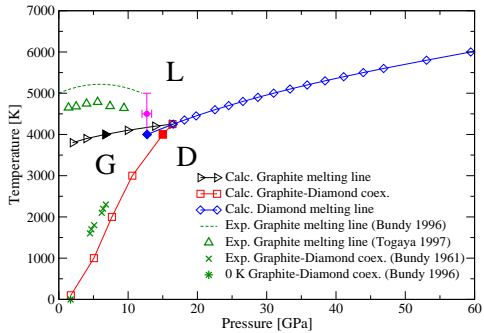}
\caption{Phase diagram of carbon up to 60 GPa. The solid right
triangle, square, and diamond are the three coexistence points
found by equating the chemical potentials at 4000~K (see text).
The open right triangles, squares, and diamonds are the calculated
coexistence points, propagated via Gibbs-Duhem integration. The
solid circle with error bars indicates the experimental estimate
for the liquid/graphite/diamond triple point~\cite{BundyC,
Glosli,vanThielReePRB}. The dashed curve is the experimental
graphite melting curve from Ref.~\cite{BundyC}. The up triangles
are graphite melting state points from Ref.~\cite{TogayaPRL}. The
crosses represent experimental estimates for graphite/diamond
coexistence from Ref.~\cite{BundyGDCL}. The asterisk represent the
theoretical graphite/diamond coexistence at zero kelvin, as
reported in Ref.~\cite{BundyC}.} \label{PD}
\end{figure}
The intersections of the chemical potential curves yield the
graphite/liquid coexistence at 6.72 $\pm$ 0.60~GPa
($\mu_\textrm{GL} = -24.21 \pm 0.10$~$k_\textrm{B}T$), and the
graphite/diamond coexistence at 15.05 $\pm$ 0.30~GPa
($\mu_\textrm{GD} = -23.01 \pm 0.03$~$k_\textrm{B}T$). The third
intersection locates a diamond/liquid coexistence at $=$~12.75
$\pm$ 0.20 GPa ($\mu_\textrm{DL} = -23.24 \pm
0.03$~$k_\textrm{B}T$). Even though both diamond and the liquid
are there metastable, the Clausius-Clapeyron integration of the
diamond melting curve can be started at the metastable coexistence
point at 4000K. Starting from the three coexistence points at
4000~K, the coexistence curves were traced by integrating the
Clausius-Clapeyron equation using the trapezoidal-rule
predictor-corrector scheme~\cite{IntegrCC}. The new value of the
coexisting $P$ at a given $T$ was taken when two iterations
differed less than 0.01 GPa, this being the size of the single
uncertainty in the calculation of d$P/$d$T$. Typically this
required 2-3 iterations.

\begin{figure}[!t]
\centering
\includegraphics[width=0.75\columnwidth,clip]{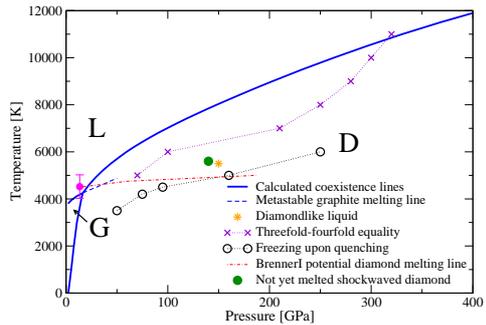}
\caption{Phase diagram of carbon at between 0 and 400 GPa. The
thick solid curves are our calculated phase boundaries. The dashed
curve is the metastable prolongation of the graphite melting
curve, from Gibbs-Duhem integration; the curve stops just before
the simulated graphite became unstable, displaying large density
fluctuations. The dashed-dotted curve departing from the
experimental estimate for the triple point (solid circle with
error bar~\cite{BundyC, Glosli,vanThielReePRB}) is the diamond
melting curve calculated in Ref.~\cite{GlosliReeJCP} with the
BrennerI potential. The solid circle is the final point of the
shock wave experiment of Ref.~\cite{Shaner} at which diamond is
not yet melted. The crosses mark the (metastable) liquid state
points with an equal fraction of three and four-fold coordinated
atoms. The circles represent state points at which the liquid
freezes. } \label{PD2}
\end{figure}

The calculated phase diagram in the $P-T$ plane is shown in
Fig.~\ref{PD} for the low pressure region, and in Fig.~\ref{PD2}
for the pressures up to 400GPa. Tab.~\ref{rho_LGD} lists the
densities of selected points on the coexistence curves. The three
coexistence curves meet in a triple point at 16.4 $\pm$ 0.7 GPa and
4250 $\pm$ 10~K.

The graphite/diamond coexistence curve agrees well with the
experimental data. In the region near the liquid/graphite/diamond
triple point that has not been directly probed in experiments, the
graphite/diamond coexistence curve bends to the right, departing
from the commonly assumed straight line. Analysis of our data
shows this is mainly due to the rapid reduction with increasing
pressure of the interplanar distance in graphite at those
premelting temperature. This causes an enhanced increase of the
density in graphite, yielding a decrease of
$\textrm{d}T/\textrm{d}P$.

\begin{table}[!t]
\begin{center}
\begin{tabular*}{0.95\columnwidth} {@{\extracolsep{\fill}}r r r r r}
  \multicolumn{4}{l}{Graphite melting curve} \\
  $P$~[GPa] & $T$~[K] & $\rho_G$  & $\rho_L$ & $\Delta h_m$\\
  \hline
\hspacetable  2.00  &    3800     &  2.134 & 1.759 & 68.8\\
  6.70     & 4000   &       2.354 & 2.098 & 66.3\\
16.4      & 4250     &      2.623   & 2.414 & 64.7\\
\hline
\multicolumn{4}{l}{Diamond melting curve}       \\
$P$~[GPa]  & $T$~[K] & $\rho_D$& $\rho_L$ & $\Delta h_m$\\
\hline
\hspacetable 16.4    &  4250   &       3.427   & 2.414 & 95.9\\
25.5     & 4750    &      3.470  & 2.607 & 111.5\\
43.9     & 5500    &   3.558       & 2.870 & 130.8\\
59.4     & 6000    &      3.629     & 3.043 & 143.9\\
99.4     & 7000    &      3.783     & 3.264 & 160.5\\
148.1     & 8000    &      3.960     & 3.485 & 164.7\\
263.2     & 10000  &      4.286     & 3.868 & 195.3\\
330.5     & 11000  &      4.230     & 4.045 & 208.1\\
408.1     & 12000  &      4.593     & 4.236 & 221.7\\
\hline
\end{tabular*}
\end{center}
\caption{Pressure ($P$), temperature ($T$), solid and liquid
densities ($\rho$), and melting enthalpy ($\Delta h_m$) along the
melting curves. Densities are expressed in $10^{3}$~kg/m$^3$,
enthalpies are in [kJ/mol].} \label{rho_LGD}
\end{table}

Table~\ref{rho_LGD} shows the melting enthalpy $\Delta h_m$ for
graphite and diamond. These are calculated as the difference in
enthalpy between the solid and the liquid at coexistence. Our
calculated melting enthalpies of graphite are significantly lower
than the values of $\sim$ 110~kJ/mol that were reported in recent
shock-heating melting experiments~\cite{BundyC,TogayaPRL}.
Nonetheless our values retain the feature of being rather constant
along the graphite melting curve. To our knowledge, no
experimental data have been reported for the melting enthalpies of
diamond. Note, that they monotonically increase with temperature.

The calculated graphite melting temperature is monotonically
increasing with pressure and is confined to small temperature
range around 4000~K.  In contrast to data inferred from
experiments it shows no maximum and is at a somewhat lower
temperature. In agreement with the experiments the coexistence
temperature is only slowly varying with pressure.  Inspection
reveals that this behavior is due to a limited variability of the
melting enthalpy, and a similar bulk modulus for liquid and
graphite yielding a volume change upon melting that is almost
constant along the melting curve.

The sign of the slope of the diamond melting curve is consistent
with the available experimental data~\cite{Shaner,BundyC} (see
Fig.~\ref{PD2}).  When compared to the diamond melting curve of
the BrennerI model~\cite{GlosliReeJCP}, the LCBOPI$^+$ diamond
melting curve has a steeper slope yielding significantly higher
temperatures for the diamond melting curve. Recently, the melting
curve of diamond in a range up to 2000~GPa has been studied by ab
initio MD simulations using density functional theory. Wang et
al.~\cite{Wang2005} determined the relative stability of the
diamond and liquid phase by evaluating the free energy of both
phases. Correa et al.~\cite{Correa2006} determined the melting
temperature using a ``two phase'' simulation method, where the
system initially consists of a liquid and a diamond structure that
are in contact. Subsequently the melting temperature is estimated
by locating the temperature at which the system spontaneously
evolves towards a liquid or a crystalline structure. In both ab
initio MD studies it was found that the diamond melting curve
shows a maximum; around 450~GPa \cite{Correa2006} or 630~GPa
\cite{Wang2005} \footnote{The difference between these two values
gives a hint on the uncertainties related to the two different
methods used for calculating coexistence, given that the DF-MD
set-up is quite similar in the two works}. Subsequent laser-shock
experiments~\cite{Brygoo2007} provided data consistent with this
observation, indicating a negative melting slope most probably in
the region of 300-500~GPa. When comparing the LCBOPI$^+$ diamond
melting curve, that monotonically increases with pressure, to the
ab initio MD results of Refs.~\cite{Wang2005,Correa2006} we see a
significant deviation from 200~GPa onwards. This might be
attributed to an incorrect description of the liquid structure at
high compression. Indeed, LCBOPI$^+$ has not been validated
against high density structures with coordination beyond four.
These are typical configuration that might become more dominant in
the pressure region beyond 200~GPa.

\section{Existence of a liquid--liquid phase transition?}
\subsection{History of the LLPT near the graphite melting line}
\label{LLPT} \textbf{Analysis of experimental data} The
possibility of a LLPT in liquid carbon was first investigated by
Korsunskaya {\em et al.}~\cite{Korsunskaya}, who analysed data on
the graphite melting curve proposed by Bundy~\cite{BundyGML},
(those data showed a maximum melting temperature at 6.5 GPa). By
fitting the data from Bundy into the original two levels model of
Kittel~\cite{Kittel} and postulating the existence of two liquids,
Korsunskaya {\em et al.} found the critical temperature $T_c$ of
the LLPT. The model is fitted with \emph{three} points on the
graphite melting curve, with the respective derivatives, and with
the heat of melting at \emph{one} selected pressure. The fitting
procedure gives an estimate for the critical pressure of $\sim
6.5$~GPa and for the critical temperature of the searched
transition at 3770~K, i.e. below the melting temperature. The
fitted value for the entropy of freezing is the same for the two
liquids, thus implying a vertical slope (d$T/$d$P$) of the
coexistence curve (in the metastable liquid region just below the
critical temperature)\\
On the basis of these results, the authors were able to calculate also
the diamond melting curve: they predicted it to have a negative slope,
in accordance with the commonly accepted interpretation of the
experiments of Alder and Christian~\cite{Alder}. Note that the slope
of the {\em graphite} melting curve, and the slope of the
diamond/graphite coexistence, as extracted from Bundy's
data~\cite{BundyGDCL,BundyGML}, together with the densities of the
phases obtained by fitting to the two levels model {\em implied} (via
Clausius-Clapeyron equation) a negative slope of the diamond melting
curve. Different values of the slopes of the graphite boundary curves,
and of the densities of the phases can yield rather different slope of
the diamond melting
curve.\\
Consistent with the slope of the fitted graphite melting curve,
the low density liquid is predicted less dense than the coexisting
graphite, and the high density liquid more dense than the
coexisting graphite. The nature of the two liquids was predicted
as follows: at low pressure graphite melts into a liquid of
neutral particles that interact predominantly through dispersion
(London) forces. Upon increasing pressure the liquid would
transform into a metallic close packed liquid. No
assumptions were made on the local structure.\\
\textbf{A semi-empirical equation of state} The modern discussion
on the LLPT for carbon, starts with the elaboration of a
semi-empirical equation of state for carbon, valid also at high
$P$ and $T$, by van Thiel and
Ree~\cite{vanThielReeHiP,vanThielReePRB}. This equation of state
was constructed on the basis of experimental data and electronic
structure calculations. It postulated the existence, in the
graphite melt, of a mixture of an $sp_2$ and an $sp_3$ liquid. By
assuming the model of {\em pseudo-binary mixture} for the
description of the mixing of the two liquids~\cite{Tanaka00}, Van
Thiel and Ree showed that fitting their empirical equation of
state to the graphite melting points of Bundy~\cite{BundyGML},
they predict a graphite melting curve that shows a maximum with a
discontinuous change of the slope, so that a first order LLPT
arises. On the other hand, if they fit their model to the data
from Ref.~\cite{Fateeva}, the predicted $T_c$ of the LLPT drops
below the melting curve and the transition between the two liquids
becomes continuous in the stable liquid region. As pointed out by
Ponyatovsky~\cite{Ponyatovsky} the expression for the mixing
energy of the two liquid as proposed by van Thiel and Ree
in~\cite{vanThielReeHiP,vanThielReePRB} involves two ambiguities.
Firstly, extrapolating the coexistence curve between the two
liquids at atmospheric pressure, the coexistence temperature would
be $T \sim 3700$~K: this would imply that the $sp_3$ liquid (and
the glass) would be more stable than the $sp_2$ at ambient
pressure up to very high temperatures, which is in disagreement
with the experimental data. Furthermore, the mixing energy is
proposed to have a linear dependence on $T$, so that, when $T
\rightarrow 0$, also the mixing energy would tend to zero, i.e. at
zero temperature the regular solution would become an ideal
solution.
This would be rather unusual.\\
\textbf{Experimental evidence from the graphite melting
  curve}\label{sec:Togaya} Using flash-heating experiments
Togaya~\cite{TogayaPRL} determined the melting line of graphite
and found a maximum in the melting curve at $P_{max} = 5.6$ GPa.
This author fitted the six experimental points with two straight
lines: with positive slope at pressures lower than $P_{max}$, with
negative slope at pressures higher than $P_{max}$. The apparent
discontinuity at the maximum would imply the presence of a triple
point graphite/ low-density-liquid (LDL) / high-density-liquid
(HDL), as a starting point of a LLPT coexistence curve.\\
\textbf{Prediction of a short-range bond-order potential}
\label{LLPTGlosli} In Ref.~\cite{Glosli} Glosli and Ree reported a
complete study of a LLPT simulated with the `BrennerI' bond-order
potential~\cite{Brenner90} in its version with torsional
interactions~\cite{Brenner91}. The authors simulated in the
canonical (NVT) ensemble several samples at increasing densities
at eight different temperatures. By measuring the pressure, they
show the familiar van der Waals loop betraying mechanical
instability over a finite density range. Using the Maxwell
equal-area construction, the authors calculated the LLPT
coexistence curve, ending in a critical point at $T = 8802$~K and
$P = 10.56$~GPa. The lowest temperature coexistence point was
calculated at  $T = 5500$~K and $P = 2.696$~GPa. The LDL/HDL
coexistence curve should meet the graphite melting curve at its
maximum, but unfortunately the BrennerI potential does not contain
non bonded interactions, thus it can describe neither bulk
graphite nor its melting curve. To overcome this deficiency, the
authors devised an ingenious perturbation method. Assuming
constant slope of the negative sloped branch of the graphite
melting curve (the authors adopted the graphite melting curve
measured by Togaya~\cite{TogayaPRL}) and fixing the
graphite/diamond/HDL triple point at a value taken from the
experimental literature, they give an estimate of the
graphite/LDL/HDL triple point, at $T = 5133$~K and $P = 1.88$~GPa.
The LDL was found to be mainly two-fold ($sp$) coordinated with a
polymeric-like structure, while the HDL was found to be a network
forming, mainly four-fold, ($sp_3$) liquid. Following the
predictions of this bond-order potential, the $sp_2$ coordinated
atoms would be completely avoided in the liquid. The authors
identified the reason in the presence of torsional interactions.
In fact, the increase in density demands an increase in structures
with higher coordination than the $sp$, which is entropically
favored at low densities. The single bonds of the $sp_3$
structures can freely rotate around the bond axis, while bonds
between $sp_2$ sites are constrained in a (almost) planar geometry
by the torsional interactions: this implies a low entropy for a
liquid dominated by $sp_2$ sites. This low entropy would
eventually destabilize the $sp_2$ sites towards the $sp_3$. To
prove this conjecture, the authors calculated two relevant
isotherms in the original version of the potential, without
torsional interactions, finding no sign of a LLPT. Since some
torsional interactions are definitely needed to mimic the double
bond reluctancy to twist, the authors concluded that the LLPT
predicted by the Brenner bond-order potential with torsion is more
realistic than its absence when torsional interactions are
switched off.\\
Tight binding calculations~\cite{TBcarbon} showed no evidence of
van der Waals loops at some of the temperatures analyzed in
Ref.~\cite{Glosli}. As Glosli and Ree note, the tight binding
model used in \cite{TBcarbon} is strictly two-center, thus the
torsional interactions {\em cannot} be described.\\
\textbf{An ab initio study of the LLPT} \label{wu} In
Ref.~\cite{Wu2002}, Wu {\em et al.} reported a series of NVT-CPMD
simulations at 6000~K from density 1.27 to 3.02~10$^3$~kg/m$^3$,
in a range where the BrennerI potential showed the first order
LLPT at the same $T$. No sign of a van der Waals loop was found:
in contrast to the BrennerI results of the previous paragraph, two
approaching series starting from the lowest and the highest
density, were found to meet smoothly at intermediate densities.
Looking for the reasons of the failure of the BrennerI potential,
the authors calculated, with the same functional used in the CPMD
simulations, the torsional energy of two model molecules. One,
(CH$_3$)$_2$C=C(CH$_3$)$_2$, was chosen so that the bond between
the two central atoms represents a double bond in a carbon
network: two $sp_2$ sites are bonded each to two $sp_3$ sites; the
peripheral hydrogens are needed to saturated the $sp_3$ atoms and
are intended to have no effect on the central bond. The second
molecule, (CH$_2$)$_2$C-C(CH$_2$)$_2$ is a portion of a completely
$sp_2$ coordinated network: in the bond-order language, the
central bond is conjugated. The two molecules were geometrically
optimized in their planar configurations and then twisted around
the central bond axis in steps of $\pi/12$. In each configuration
the electronic wave function was optimized, without further
relaxations, to give the total energy, that was compared to the
planar configuration total energy. The difference is the torsional
energy. The DF calculations found a surprising picture: while the
double bond torsional energy was only slightly overestimated by
the BrennerI potential at intermediate angles, the DF torsional
energy for the conjugated bond showed a completely different
scenario compared to the classical prediction. It shows a maximum
at $\pi/4$, while the planar and orthogonal configuration have
basically the same energy. For the BrennerI potential, the
torsional energy in this conjugated configuration is monotonically
increasing with the torsion angle, just as for the double bond
configuration. On average, considering that the conjugated
configuration would be characteristic of a mainly $sp_2$
coordinated liquid, the torsional interactions are enormously
overestimated by the classical potential. As a further
illustration, the authors tried to lower torsional energy of the
conjugated bond in the classical potential, by tuning the proper
parameter, and found a much less pronounced LLPT. Note that the
functional form of the torsional interactions for the BrennerI
potential {\em cannot} reproduce the DF data mentioned here. Wu
{\em et al.} concluded that ``[the] Brenner potential
significantly overestimates the torsional barrier of a chemical
bond between two- and three-center-coordinated carbon atoms due to
the inability of the potential to describe lone pair electrons'';
and: ``[the] Brenner potential parameters derived from isolated
hydrocarbon molecules and used in the literature to simulate
various carbon systems may not be adequate to use for condensed
phases, especially so in the presence of lone pair electrons''. In
the next section we show that the conclusion of Wu {\em et al.} is
not necessarily true for all BOPs; indeed, LCBOPI$^+$, the carbon
bond-order potential proposed by Los and Fasolino (see section
\ref{LosFasolino}), includes a definition of the torsional
interaction which is able to reproduce relevant features of liquid
carbon, as they are described by DF-MD.

\subsection{Ruling out the LLPT in the stable liquid region via LCBOP}
\label{sec:nollpt} We have already indicated that the
change of the structure of the liquid along the graphite and
diamond melting curve is related to the slope of the melting
curve.  More importantly, it plays also a crucial role in the
nucleation of diamond in liquid carbon. The latter will be further
discussed in the next section.

The calculated melting curves of the LCBOPI$^+$ model for carbon
up to 400 GPa provide strong evidence that there is no LLPT in the
stable liquid phase. One indication is the smoothness of the
slopes of the melting curves. A further argument lies in the
structure of the liquid near freezing. Below we discuss this
in more detail.

The calculated phase diagram (Figs.~\ref{PD} and \ref{PD2}) does
not show the sharp maximum in the graphite melting line that was
inferred from the calculated first-order LLPT for the BrennerI
bond-order potential\cite{Glosli}. As we mentioned in the previous
section, subsequent DF-MD simulations of liquid
carbon~\cite{Wu2002,Ghiringhelli04} indicate that the BrennerI
LLPT is spurious: it originates from an inadequate description of
the torsional contribution to the interactions. We have extended
the calculation of the graphite melting curve of LCBOPI$^+$
towards higher pressures into the region where both graphite and
the liquid are metastable with respect to diamond. It is plotted
as a dashed curve in Fig.~\ref{PD2} that shows the same trend as
at lower pressures. Hence, the calculated slope of the graphite
melting curve is incompatible with the existence of a LLPT in this
region of the carbon phase diagram.

\begin{figure}[!t]
\centering
\includegraphics[width=0.75\columnwidth,clip]{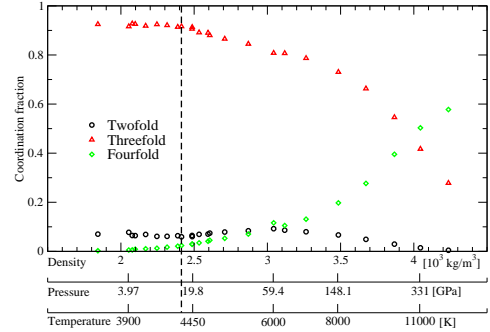}
\caption{Coordination fraction of the liquid along the melting
curve of carbon. The melting curves are unimodal, thus fixing
$\rho$ providing one-to-one relation among $\rho$, $P$ and $T$.
The scale for $\rho$ is chosen to be linear.  The dashed curve is
the liquid/graphite/diamond triple point. On the left hand side of
the triple point, the liquid coexists with graphite, while on the
right hand side it coexists with diamond.} \label{coordcoex}
\end{figure}

In order to further analyze the nature of the liquid, we
determined several structural properties of the liquid near the
melting curve where we also explored the diamond melting curve.
Fig.~\ref{coordcoex} shows the coordination fraction in the liquid
along the coexistence curves up to 400~GPa, as function of
temperature, pressure and density, with a linear scale in density.
The dashed curve is the calculated graphite/diamond/liquid triple
point. Along the graphite melting curve, the three-fold and
two-fold coordination fractions remain rather constant, with the
four-fold coordination slightly increasing to account for the
increase in density.  Along the diamond melting curve the
three-fold coordinated atoms are gradually replaced by four-fold
coordinated atoms. However, only at (3.9~10$^3$~kg/m$^3$, 300~GPa,
and 10500~K) the liquid has an equal fraction of three-fold and
four-fold coordinated atoms.  The change of dominant coordination
is rather smooth. Moreover, we have verified that it is fully
reversible showing no sign of hysteresis in the region around the
swapping of dominant coordination. We note, that these results
contradict the generally assumed picture (see e.g.
Ref.~\cite{vanThielReePRB}) that diamond melts into a four-fold
coordinated liquid. Our calculations suggest that up to
$\sim$300~GPa the three-fold coordination dominates.

\begin{figure}[!t]
\begin{center}
\includegraphics[width=0.9\columnwidth,clip]{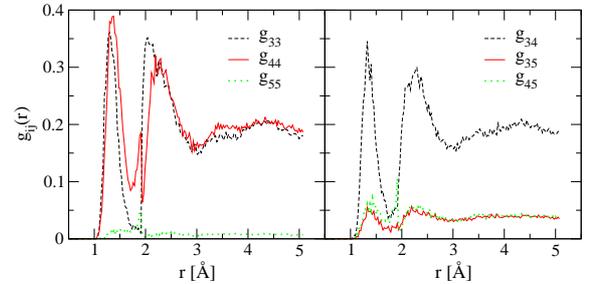}
\caption{Partial distribution functions $g_{ij}$ of the liquid at
the calculated coexistence with diamond, at 10500~K and
$\sim$~300~GPa, when three- and four-fold atoms are equally
present. The left panel is for the diagonal contributions (i.e.
for $i=j$), while the right panel is for the cross correlations
(i.e. for $i \neq j$).} \label{PRDF}
\end{center}
\end{figure}

The interrelation between three and four-fold sites, was further
investigated calculating the partial radial distribution functions
($g_{ij}(r)$) of the liquid at 300~GPa, and 10500~K. Partial
radial distribution functions are defined as the probability of
finding a $j$-fold site at a distance $r$ from a $i$-fold site;
the total radial distribution function $g$ is recovered by:
$g=\sum_{i} g_{ii} + 2\sum_{i \neq j} g_{ij}$. We show the results
in Fig.~\ref{PRDF}; we focus on the three predominant curves,
describing the pair correlations between three-fold atoms
($g_{33}$), between four-fold atoms ($g_{44}$), and the cross pair
correlation between three- and four-fold sites ($g_{34}$).
Disregarding the rather pronounced minimum in correspondence of
the dip around 2~\AA~ of the $g_{33}$ and the $g_{34}$, the
similarity of three curves at all distances $r$ is striking. The
two sites are almost undistinguishable: in case of a tendency
towards a phase transition, one would expect some segregation of
the two structures. In contrast, looking at distances within the
first neighbours shell, a three-fold site seems to bond
indifferently to a three- or a four-fold site, and viceversa.
Furthermore, the partial structures up to the third, quite
pronounced, peak at $\sim$~4.5~\AA, are almost the same for these
three partial radial distribution functions.

We determined the properties of the metastable liquid in the
stable diamond region.  Fig.~\ref{PD2} shows the liquid $P-T$
state points (crosses) that exhibit an equal number of three and
four-fold coordinated atoms.  It ranges from the high-pressure
high-temperature region where the liquid is thermodynamically
stable down into the diamond region, where the liquid is
metastable for the LCBOP. The circles indicate state points in
which the LCBOP liquid freezes in the simulation. Enclosed by the
two set of points lies what we baptized diamond-like liquid. This
is a mainly four-fold coordinated liquid with a rather pronounced
diamond-like structure in the first coordination shell and was
discussed in \cite{Ghiringhelli04}. This suggests that a
(meta)stable liquid with a dominantly four-fold coordination may
only exist for pressures beyond $\approx$ 100~GPa and could imply
that the freezing of liquid into a diamond structure might be
severely hindered for a large range of pressures beyond the
graphite/diamond/liquid triple point.  In
Ref.~\cite{Ghiringhelli04} it is also pointed out that at 6000~K
the equation of state shows a change of slope around the
transition to the four-fold liquid. At even lower temperature this
feature becomes more and more evident, but for temperatures lower
than $\sim$4500~K the liquid freezes into a mainly four-fold
coordinated amorphous structure.  This observation is consistent
with quenching MD simulations~\cite{MarksEDIPRev,MarksComp} to
obtain the tetrahedral amorphous carbon. In those simulations a
mainly three-fold liquid freezes into an almost completely
four-fold amorphous.

Recent fully ab-initio study of the diamond melting line
\cite{Wang2005,Correa2006} predicted a maximum at pressures beyond
the maximum pressure (400 GPa) we explored with our potential. The
maximum implies a liquid denser than diamond at pressure higher
than the pressure at which the maximum appears. The authors of
both works analyzed the structure of the liquid around this
maximum, finding no sign of abrupt change in density and/or
coordination. This points towards excluding a LLPT between a
four-fold and a higher-fold coordinate liquid; rather, a smooth
transformation towards a denser liquid is always observed.

\section{Diamond nucleation}\label{dianucl}
Our knowledge of the phase diagram of ``LCBOPI$^+$ carbon'' allows
us to identify the regions of the phase diagram where diamond
nucleation may occur. We studied the homogeneous nucleation of
diamond from bulk liquid, by computing the steady-state nucleation
rate and analyzing the pathways to diamond formation. On the basis
of our calculations, we speculate that the mechanism for
nucleation control is relevant for crystallization in many
network-forming liquids, and also estimate the conditions under
which homogeneous diamond nucleation is likely in carbon-rich
stars and planets such as Uranus and Neptune.

{\bf Steady-state nucleation rate} Most liquids can be cooled
considerably below their  equilibrium freezing point before
crystals start to form spontaneously in the bulk. This is caused
by the fact that microscopic crystallites are thermodynamically
less stable than  the bulk solid. Spontaneous crystal growth can
only proceed when, due to some rare fluctuation, one or more
micro-crystallites exceed a critical size (the ``critical
cluster''): this phenomenon is called homogeneous nucleation. An
estimate of the free-energy barrier the system has to cross in
order to form critical clusters and of the rate at which those
clusters form in a bulk super-cooled liquid, can be obtained from
Classical Nucleation Theory (CNT)~\cite{kelton}. CNT assumes that
$\Delta G(n)$, the Gibbs free-energy difference between the
metastable liquid containing an {\it n}-particle crystal cluster
and the pure liquid, is given by
\begin{equation}
\Delta G(n)= S(n)\gamma-n|\Delta\mu|, \label{CNTbarrier-N}
\end{equation}
where $S(n)$ is the area of the interface between an n-particle
crystallite and the metastable liquid, $\gamma$ is the
liquid-solid surface free-energy per unit area, and $\Delta \mu$
the difference in chemical potential between the solid and the
super-cooled liquid. The surface area $S(n)$ is proportional to
$c(n/\rho_{S})^{2/3}$, where the factor $c$ depends on the shape
and the geometry of the cluster  (e.g. $c=16\pi/3$ for a spherical
cluster).

The top of the free-energy barrier  $\Delta G^{*}$ to grow the
crystalline critical cluster is then given by
\begin{equation}
\Delta G^{*} = c \frac{\gamma^{3}}{\rho_{S}^{2}|\Delta \mu|^{2}} ,
\label{barrier}
\end{equation}
where $\rho_{S}$ is the number density of the stable phase and $c$
indicates the geometrical properties of the growing cluster. From
our simulations, we can only determine the product $c\gamma$: it
is this quantity and the degree of super-saturation ($\Delta
\mu$), that are needed to compute the top of the free-energy
barrier, and hence the nucleation rate.

CNT relates $R$, the steady-state nucleation rate, i.e. the number
of crystal clusters that form per second per cubic meter, to
$\Delta G^{*}$, the height of the free-energy barrier that has to
be crossed to nucleate the critical crystal $n^*$:
\begin{eqnarray}
\label{nucleationrate} R^{CNT} = \kappa  e^{-\beta \Delta G(n^*)},
\end{eqnarray}
where $\Delta G^{*}$ is  the top of the free-energy barrier and
$\kappa$ is the kinetic prefactor. The kinetic prefactor term is
defined as
\begin{equation}
\label{prefactor} \kappa = \rho_L k_{\rm +,n^*} Z
\end{equation}
where $\rho_L$ is the liquid number density, $k_{\rm +,n^*}$ the
attachment rate of single particles to a spherical crystalline
cluster $ k_{\rm +,n^*} = \left( 24 D (n^*)^{2/3} \right) /
\lambda^2 $, with $D/\lambda^2$  proportional to the jump
frequency ($\lambda$ being the atomic jump distance) and $Z =
\sqrt{|\Delta \mu| / (6\pi k_BT n^*)}$ the so-called Zeldovitch
factor. As the nucleation rate depends exponentially on $\Delta
G^{*}$, a doubling of $\gamma$ may change the nucleation rate by
many orders of magnitude.

Because of the extreme conditions under which homogeneous diamond
nucleation takes place, there have been no quantitative
experimental studies to determine its rate. Moreover, there exist
no numerical estimates of $\Delta \mu$ and $\gamma$ for diamond in
super-cooled liquid carbon. Hence, it was thus far impossible to
make even an order-of-magnitude estimate of the rate of diamond
nucleation.

{\bf Results} We simulate a 2744 particles bulk liquid carbon
using periodic boundary conditions, with a  cubic box whose edge
is 18\AA. We make it metastable by undercooling it at constant
pressure at two different state points,  $A \{P=85$~GPa,
$T=5000$~K$\}$ and $B \{ P=30$~GPa, $T=3750$~K$\}$. At both state
points, the liquid is super-cooled by $(T_{m}-T)/T_{m} \approx$
25$\%$ below the diamond melting curve, $T_{m}$ being the melting
temperature $T_{m}^{A}=$~$6600$~K and $T_{m}^{B}=5000$~K,
respectively.

We evaluate $\Delta \mu$ by thermodynamic integration at constant
pressure from the melting point ($\beta_M=1/k_BT_M$)
\begin{equation}
\label{deltamu} \Delta (\beta_i \mu) =
\int_{\beta_{M}}^{\beta_{i}} \langle  [h_S(\beta) - h_L (\beta) ]
\rangle_P d\beta
\end{equation}
where $\beta_i = 1/k_B T_i, \{i=A,B\}$, and $h$ is the
enthalpy per particle of the solid and liquid phase, respectively.
We then find: $|\Delta \mu_{A}/k_BT|=0.60$ and $|\Delta
\mu_{B}/k_BT|=0.77$.

In recent years several authors have been developing methods for
studying homogeneous nucleation from the bulk and detecting solid
particles within the metastable
liquid~\cite{tenwolde,auer,Valeriani,Valeriani-thesis}. Our study
on diamond nucleation is based on these works, but it requires
various adaptations due to the specificity of the carbon covalent
bond~\cite{Ghiringhelli-thesis}. We have already shown that liquid
carbon is rather structured below its freezing curve. This leads
to the need of building a "strict" definition of ``crystallinity''
of a particle, in order to avoid an overestimation of the number
of solid particles in the system.

In order to compute the nucleation free energy, we use the biggest
crystal cluster $n$ as a local order parameter to quantify the
transformation from the liquid to the solid. To identify
solid-like particles, we analyze the local environment of a
particle using a criterion based on a spherical-harmonics
expansion of the local bond order. In practice, the present
bond-order parameter is based on rotational invariants constructed
out of rank three spherical harmonics ($Y_{3m}$). This choice
allows us to identify the tetragonal symmetry of the diamond
structure, as already described in
Ref.~\cite{Ghiringhelli-thesis,Valeriani-thesis}, and it is also
perfectly suited to find particles in a graphite-like environment.
Our choice of odd-order of spherical harmonics is due to the fact
that both diamond and graphite lattices have odd symmetry upon
inversion of coordinates.

In order to define the local order parameter, we start with computing
\begin{equation}
\label{eq:q3} q_{3,m}(i) = \frac{1}{Z_i} \sum_{j \neq i}
S^{down}(r_{ij}) \; Y_{3m}(\hat{\textbf{r}}_{ij})
\end{equation}
where the sum extends over all neighbors of particle $i$ and over
all values of $m$.  $Z_i$ is the fractional number of neighbours and
$S^{down}(r_{ij})$ is a smooth cut-off
function, introduced in the context of
$LCBOPI^{+}$\cite{Ghiringhelli-thesis} (see also Section II.A in
\cite{LCBOPII-I}).

By properly normalizing Eq.~\ref{eq:q3}, we get
\begin{equation}
\label{eq:q3_1} q_{3,m}^{'}(i) =
\frac{q_{3,m}(i)}{(\sum_{m=-l}^{l}q_{3,m}(i)  \cdot q_{3,m}^{*}(i)
)^{1/2}},
\end{equation}
being $q_{3,m}^{*}$ the complex conjugate of $q_{3,m}$.

Next we define the dot product between the normalized function
$q_{3,m}^{'}$ of particle $i$ and the same function computed for
each of its first neighbors, $d_3(i,j)$, and sum them  up over all
the $m$ values:
\begin{equation}
\label{eq:q3_2} d_{3}(i,j) = \sum_{m=-l}^{l} q_{3,m}^{'}(i) \cdot
q_{3,m}^{'  *}(j) S^{down}(r_{ij}).
\end{equation}
$d_{3}(i,j)$ is a real number defined between -1 and 1: it assumes
the value of -1 when computed for both graphite and diamond ideal
structures.

Two neighboring particles $i$ and $j$ are considered to be
connected whenever $d_{3}(i,j) \leq d_c=-0.87$. This value
satisfactorily splits the distributions of solid particles
belonging to a thermalized lattice and liquid particles as found in a
liquid. The histograms that led us to this choice are
thoroughly discussed in Refs.~\cite{Ghiringhelli-thesis,Valeriani-thesis}.
By counting the total number of connections ($n_{con}$) and plotting the
probability distribution of $n_{con}$, we define a threshold for
the number of  connections needed to neatly distinguish between a
liquid-like and a solid-like environment: we assume that whenever
$n_{con}>n_{con}^c=3$ a particle is solid-like. At this stage, we
do not specify any nature of the particle's crystallinity, whether
diamond-like or graphite-like. By means of a cluster algorithm we
then define all the solid-like AND connected particles as
belonging to the same crystal cluster. After computing the size of
each cluster, we use the size of the biggest cluster as the
order parameter which describes the phase transition~\cite{tenWoldeTH}.

Once properly identified the biggest crystalline cluster  in the
system, we use the umbrella sampling technique~\cite{UmbSam} to
measure the free-energy barrier $\Delta G^{*}$ to form a critical
cluster at state point $A$. In order to better equilibrate the
growing clusters, we implement a ``parallel tempering'' algorithm
similar to the one described in Ref.~\cite{ParTempAuerFrenkel}. We
obtain that, at state point $A$, $\Delta G^{*}_{A}$ is around 25
$k_{B}T$ for a critical cluster size of $n_A=110$. By fitting the
initial slope of Eq. \ref{CNTbarrier-N} to a polynomial function
assuming a spherical growing cluster, while imposing the value of
the correspondent super-saturation ($\beta \Delta \mu=$0.60), the
inter-facial free energy is $\gamma_{A} = 0.27 k_{B}T/$\AA$^2
\simeq 1.86$ J m$^{-2}$. The same value of $\gamma_{A}$ is
obtained from the top of the free-energy barrier assuming a
spherical cluster shape (Eq.~\ref{barrier}) ($\rho_{S}=0.191$
\AA$^{-3}$). We underline the fact that at the chosen
thermodynamic conditions, there are no finite size effects, caused
by  {\it spurious} interaction of the critical cluster with its
own periodically repeated image.

By knowing the inter-facial free energy in $A$, and assuming the
validity of CNT, we estimate the crystal nucleation rate by means
of Eq.~\ref{nucleationrate}, where we use Eq.~\ref{prefactor} to
compute the kinetic pre-factor (the atomic jump distance $\lambda$
being of the order of the diamond bond distance, $1.54$\AA):
$R_A^{CNT}\sim O(10^{30})$~s$^{-1}$m$^{-3}$.

We also use Forward-Flux Sampling (FFS), a relatively recent rare
events technique useful to compute the nucleation rate and to
study the pathways to nucleation~\cite{FFS,FFS2,Valeriani-thesis},
and we measure the crystal nucleation rate at state point $A$. FFS
yields an estimate for the nucleation rate that is three orders of
magnitude higher than the one estimated by means of
Eq.~\ref{nucleationrate}. Whilst such a discrepancy seems large,
it need not be significant because nucleation rates are extremely
sensitive to small errors in the calculation of the nucleation
barrier. Two possible reasons for this discrepancy are: 1) if we
consider that the standard deviation corresponding to $\gamma$ is
around $10\%$ of its measured value, we conclude that the
nucleation rate is $O(10^{30\pm3})$~s$^{-1}$m$^{-3}$; 2) another
source of error can be the poor statistics when  computing  the
nucleation rate from molten carbon by means of FFS. This is due to
the time consuming calculations of the interaction potential: in
our study we are in fact forced to base our results on $O(10)$
independent nucleation events. Fig.~\ref{cluster85GPa} shows a
typical critical cluster at state point $A$ obtained in the FFS
simulations:
\begin{figure}[h]
\begin{center}
\mbox{\includegraphics[clip=true,width=0.4\columnwidth]{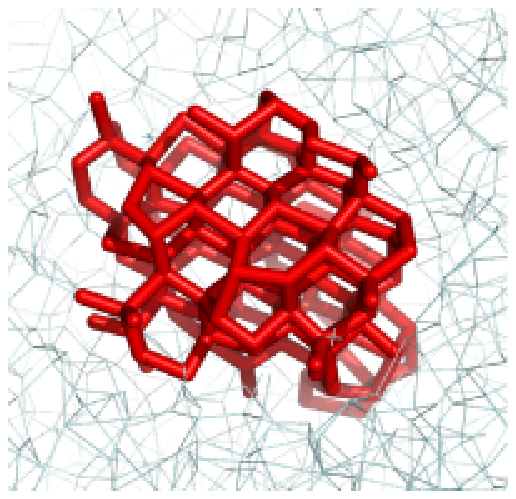}
\hspace{0.5cm}
\includegraphics[clip=true,width=0.4\columnwidth]{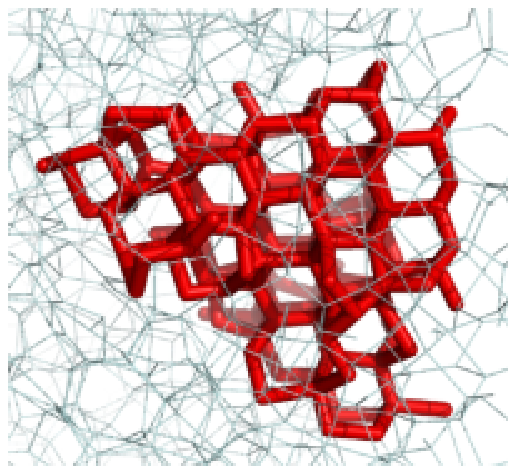}}
\caption{{\small Two different views of the biggest cluster at
state point $A$ containing around 110 particles, surrounded by
mainly 4-fold coordinated liquid particles.\label{cluster85GPa}}}
\end{center}
\end{figure}
it contains around 110 particles, and it is surrounded by mainly
4-fold coordinated liquid particles. The picture shows two
different views of the same cluster: it appears evident that all
particles within the bulk are diamond-like, whereas the particles
belonging to the outer surface are less connected but still mainly
3-4 fold coordinated.

We then attempt to compute the nucleation rate at state point $B$
by means of FFS and, even in rather long simulations, we cannot
observe the formation of any crystal cluster containing more than
75 particles. Hence, these calculations suggest that the
nucleation rate at state point $B$ measured by means of FFS is
around zero.
\begin{figure}[h]
\begin{center}
\includegraphics[clip=true,width=0.7\columnwidth]{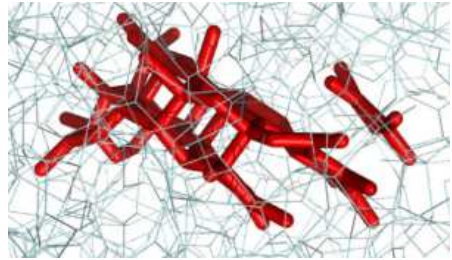}
\caption{{\small Typical snapshot of a crystalline cluster of
$\sim 75$ particles obtained at state point $B$, surrounded by
mainly 3-fold coordinated liquid particles.}}\label{fig:gradia}
\end{center}
\end{figure}
Figure~\ref{fig:gradia} shows a 75 particles cluster containing
3-fold coordinated surface particles surrounding the 4-fold
coordinated bulk particles, while embedded in a 2-3 fold
coordinated liquid.

As we are unable to grow critical nuclei with FFS, we assume that
a system of 2744 particles is too small to accommodate a spherical
critical cluster. According to Classical Nucleation Theory
(CNT)~\cite{kelton}, the crystal nucleation rate depends
exponentially on the height of the free-energy barrier (see
Eq.~\ref{nucleationrate}). The latter is a function of the
inter-facial free energy ($\gamma$) cube and inversely
proportional to the super-saturation ($\Delta \mu$) square. Since
the super-saturation is quite similar in both state points, the
failure of the system to nucleate suggests that the inter-facial
free energy should play a major role. In order to estimate the
free-energy barrier in state point $B$, as we know the solid
number density ($\rho_{S}=0.177$~\AA$^{-3}$) and the chemical
potential difference between the liquid and the solid ($\beta
\Delta \mu_{B}=$0.77), we only need to calculate the inter-facial
free energy $\gamma$. Thus, in what follows, we focus on methods
to estimate  $\gamma$ at state point $B$.

As a spherical critical cluster does not fit in our
simulation box, we prepare a rod-like crystal in a system with a
slab geometry: this is a flattened box containing around $4000$
particles, with lateral dimensions that are some four times larger
than its height. The crystal rod is oriented perpendicular to the
plane of the slab, it spans the height of the simulation box and
is continued periodically. The cross section of this crystal rod
is initially lozenge shaped, such that its [111]-faces are in
contact with the liquid. The [111]-planes are the most stable ones
for the diamond lattice. In fact, macroscopic natural diamonds
have often an octahedral shape, with eight [111]-exposed surfaces.
Indeed, we find stable [111] surfaces in all but the smallest
studied diamond clusters.

At state point $A$ clusters grow by the addition of particles
to the surface made of mainly [111]-planes.
Note that when graphite and diamond structures compete at state
point $B$ (as shown in Fig.~\ref{fig:gradia}), the [0001]-graphite
sheets transform into [111] diamond planes.
\begin{figure}[h]
\begin{center}
\includegraphics[clip=true,width=0.6\columnwidth]{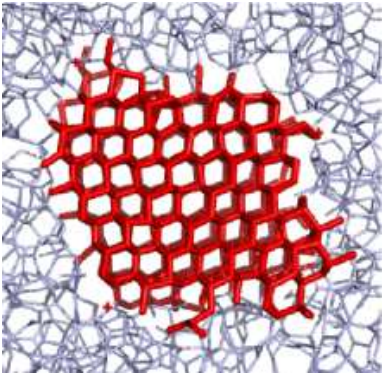}
\caption{{\small Top view of a rectangular parallelepiped formed
by 4 [111]-faces and 2 [1-10]-lozenge bases with an acute angle
$\theta$=70.52 degrees thermalized at state point $B$. \label{lozenge1-2}}}
\end{center}
\end{figure}
Fig.~\ref{lozenge1-2}  represents the top view at state point $B$
of a rod-like crystal, formed by 4 [111]-faces and 2 bases as
[1-10]-lozenge with the acute angles of $\theta$=70.52 degrees. We
then rewrite Eq.~\ref{CNTbarrier-N} for a rectangular
parallelepiped having 4 faces and 2 lozenge-shaped basis
\begin{equation}
\Delta G =  4 \sqrt{\frac{h}{\rho \sin\theta}} \gamma^{l} n^{1/2}
-|\Delta \mu| n, \label{loze}
\end{equation}
where h is the slab's height. We then use Umbrella Sampling to
compute the initial slope of the free-energy barrier. As $h=10$
\AA, we obtain from fitting Eq.~\ref{loze} that the inter-facial
free energy for the lozenge-shaped cluster is $\gamma^{l}_{B} =
0.91 k_{B}T/$\AA$^2 \simeq 4.70$ J m$^{-2}$. At the same time,
computing the inter-facial free energy of the rod-like crystal at
state point $A$ gives $\gamma^{l}_{A} = 0.37 k_{B}T/$\AA$^2 \simeq
2.55$ J m$^{-2}$, considering the same slab's height and the same
angle $\theta$.

Now that we have estimates for the inter-facial free energies of
the lozenge-shaped clusters at both state points $A$ and $B$, we
can estimate the ratio between them and find that $c \gamma_{B}/ c
\gamma_{A} = \gamma_{B}/ \gamma_{A} \simeq 2.5$. As we compare
clusters having the same shape, this ratio is presumably not very
sensitive to the precise (and, a priori unknown) shape of the
cluster shape. As the surface free energy at state point B is
appreciably higher than at state point A, the early stages of
crystal formation at point $B$ are strongly suppressed by the
inter-facial free-energy term. Since we know $\gamma_{A}$
(referred to a hypothetical spherical cluster)  and the ratio
between the two $\gamma$'s, we can infer that the ``effective''
$\gamma_{B} = 0.68 k_{B}T/$\AA$^2 \simeq$ 3.50 J m$^{-2}$.

This value of the surface free energy is so large that we would
indeed have needed a much larger system in order to accommodate
the critical cluster  at the state $B$ thermodynamic conditions.
>From Eq.~\ref{loze}, we calculate the critical cluster size for
the lozenge-shaped parallelepiped $n^{*}_{2\mathrm{D}}$ and use it
to estimate the size of a critical spherical cluster
$n^{*}_{3\mathrm{D}}$ in $B$:
\begin{equation}
n^{*}_{2\mathrm{D}} = \frac{4 h}{\rho_{S} \sin \theta}
\frac{(\gamma)^{2}}{ (\Delta \mu)^2 }. \label{Nloz}
\end{equation}
At state point $B$ we find $B$ $n^{*}_{2\mathrm{D}} \sim$330
particles. Expressing $n^{*}_{3\mathrm{D}}$ as a function of the
lozenge-shaped parallelepiped one, we get
\begin{equation}
n^{*}_{3\mathrm{D}} = \frac{8}{3} \pi \frac{\gamma \sin
\theta}{\rho_{S} \Delta \mu h} \times n^{*}_{2\mathrm{D}},
\label{NspfromNloz}
\end{equation}
where $\rho_{S}$ is the solid number density $\rho_{B}=0.17$~\AA$
^{-3}$, $|\Delta \mu_{B}/k_BT|=0.77$, and $h$ the height of the
slab (10 \AA). Thus, $n^{*}_{3\mathrm{D}}\sim$ 1700 particles at
state point $B$. To guarantee that the critical cluster does not
interact with its own periodic images, its radius should always be
less than 25\% of the box diameter $L$. A spherical cluster with a
radius of 0.25 $L$ occupies $\sim$ 7 \% of the volume of the box
and, as the solid is denser than the liquid, it  contains about 10
\% of the total number of particles ($N\approx 17000$). Such a
large system size is beyond our present computational capacity. In
contrast, in the slab geometry we find that the free energy of a
lozenge-shaped crystal goes through a maximum at a size of $\sim$
330 particles, which is much less than the system size (4000
particles).

As $\Delta\mu$ and $\rho_{B}$ are known, we can now use CNT to
estimate $\Delta G^{*}$ in state point $B$. It turns out that,
mainly because $\gamma_{B}$ is 2.5 times larger than $\gamma_{A}$,
the nucleation barrier in $B$ is more than ten times higher than
in point $A$, and the nucleation rate is $R_{B} \sim
10^{-80}$~s$^{-1}$m$^{-3}$.

To understand the microscopic origin for the large difference in
nucleation rates in state points $A$ and $B$, it is useful to
compare the local structure of the liquid phase in both state
points.   As discussed in section~\ref{LLPT} above (see
also~\cite{Ghiringhelli04,Ghiringhelli05JPCP}), liquid carbon is
mainly 4-fold coordinated at state point $A$ ($20\%$ 3-fold and
$80\%$ 4-fold ), whereas at the lower temperatures and pressures
of point $B$, the coordination in the liquid resembles that of
graphite and is mainly 3-fold coordinated ( $5\%$ 2-fold, $85\%$
3-fold and $10\%$ 4-fold ).

\begin{figure}[h!]
\begin{center}
\includegraphics[width=0.95\columnwidth,clip]{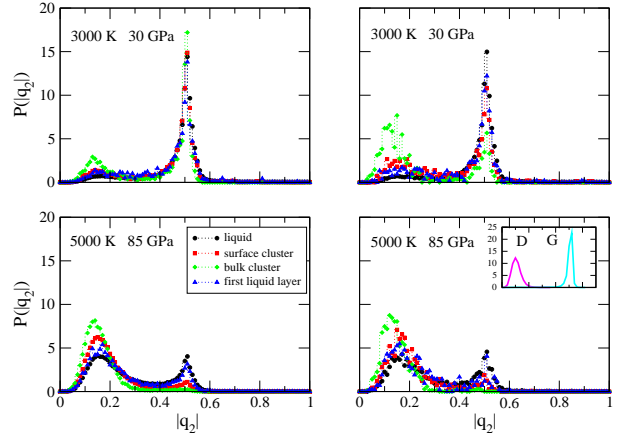}
\caption{{\small The top-left represents clusters of $\sim$ 20 and
the top-right clusters of $\sim$ 75 at state point $B$, whereas
the bottom-left clusters of $\sim$ 20 and the bottom-right
clusters of $\sim$ 250 particles at state point $A$. The used code
is: circles=liquid particles, squares=particles belonging to the
biggest cluster's surface, diamonds=particles within the bulk
cluster, triangles=particles belonging to the first liquid layer
surrounding the biggest cluster. The inset shows the $|q_{2}|$
probability distribution for an equilibrated bulk diamond (D)
(left-hand side) and graphite (G) (right-hand
side).\label{q2mod}}}
\end{center}
\end{figure}

We can analyze the structure of the crystalline clusters that form
in the supersaturated liquid carbon and distinguish graphite-like
from diamond-like particles. In an $\it{a \ posteriori}$ analysis,
we use a different order parameter function of the order two
spherical harmonics, and particularly sensitive to the graphite
planar geometry. $q_{2m}(i)$ is the linear combination of
spherical harmonics computed for each particle $i$
\begin{equation}
\label{eq:q2m} q_{2m}(i) = \frac{1}{Z_i} \sum_{j \neq i}
S^{down}(r_{ij}) \; Y_{2m}(\hat{\textbf{r}}_{ij})
\end{equation}
where the sum extends over all neighbors of particle $i$. We then
sum over all the $m$ values and calculate the modulus, $|q_{2}|$.
The  $|q_{2}|$ probability distribution for both $A$ and  $B$ is
represented in Fig.~\ref{q2mod}.
Figure~\ref{q2mod} depicts the features of both the smallest
($\sim$ 20 in both state points $A$ and $B$) and the biggest
clusters ($\sim$ 250 in $A$ and $\sim$ 75 in $B$). We  also
distinguish among: liquid-like particles (circles), particles
belonging to the surface of the largest cluster (squares),
particles inside the bulk cluster (diamonds) and particles
belonging to the first liquid layer surrounding the largest
cluster (triangles). According to our definition, particles
belonging to the surface of the cluster are those {\it connected}
to  solid-like particles, but not solid-like themselves.
Concerning particles belonging to the first liquid layer
surrounding the cluster, they usually display the same behaviour
as the ones belonging to the cluster surface, which is not
surprising in view of the uncertainty in distinguishing a
surface-particle from a first-liquid-layer particle. To  neatly
distinguish between diamond-like or graphite-like environment, we
use as a reference the $|q_{2}|$ probability distribution for both
bulk diamond (D) and graphite (G) (inset of Fig.~\ref{q2mod}).

At state point $A$, it is clear that bulk particles belonging to
small clusters (bottom-left side) and big clusters (bottom-right
side) are mainly diamond-like, as well as particles belonging to
the surface of the clusters. In contrast, at state point $B$ bulk
particles belonging to small clusters (top-left side) show both
graphite-like and diamond-like finger-prints. By visual
inspection, we note that when clusters grow larger (around 75
particles), particles at the surface tend to be mainly 3-fold
coordinated, whereas bulk particles stay 4-fold coordinated, as
shown in the top-right side of Fig.~\ref{q2mod}. The destabilizing
effect of the graphitic liquid on the diamond clusters is most
pronounced for small clusters (large surface-to-volume ratio):
clusters containing less than $25$ particles tend to be graphitic
in structure, clusters containing up to $60$ particles show a
mixed graphite-diamond structure (see Fig.~\ref{fig:gradia}). It
appears that  the unusual surface structure of the diamond cluster
is an indication of the poor match between a diamond lattice and a
3-fold coordinated liquid.

\subsection{Consequences for other network forming liquids, carbon-rich stars, Uranus and Neptune}

As discussed in the introduction, there are many network-forming
liquids that, upon changing pressure and temperature, undergo
profound structural changes or even
LLPT~\cite{Mishima98,KatayamaN,liqliq3}. Interestingly, our
simulations show that the ease of homogeneous crystal nucleation
at constant super-saturation from one-and-the-same meta-stable
liquid can be tuned by changing its pressure, and thereby its
local structure.

Pressures and temperatures that we investigate for the diamond
nucleation are in practice impossible to reach in experiments.
However, such conditions are likely to be found in several
extraterrestrial ``laboratories''. Homogeneous nucleation of
diamond may have taken place in the atmosphere of carbon-rich
binary stellar systems comprising the so-called carbon stars and
white dwarfs~\cite{whitedwarf1,whitedwarf2}. Closer to home, it has
been suggested that diamonds could also have formed in the
carbon-rich middle layer of Uranus and
Neptune~\cite{Ross,uranus1,uranus3} where, due to the high
pressure and temperature, the relatively abundant CH$_4$ would
decompose into its atomic components. In fact, experiments on
methane laser-heated in diamond anvil cells~\cite{uranus5}
found evidence for diamond production. {\it Ab
initio} simulations~\cite{uranus4} also found that hot, compressed
methane will dissociate to form diamond. Yet, there is a large
discrepancy between the estimates of the pressures (and thus depth
in the planet interior) at which the diamond formation would take
place. The laser-heating experiments~\cite{uranus5} suggested
diamond formation at pressure as low as 10--20 GPa (at 2000--3000
K), whereas the {\it ab initio} simulations~\cite{uranus4} found
dissociation of methane, but synthesis of short alkane-chains at
$\sim 100$ GPa and diamond at pressures not lower than 300 GPa
(note that simulations were carried out at 4000--5000 K).\\ The
present work allows us to make a rough estimate of the conditions
that are necessary to yield appreciable diamond nucleation on
astronomical timescales.

In this context, it is crucial to note that neither carbon stars
nor carbon-rich planets consist of pure carbon. In practice, the
carbon concentration may be as high as  $\sim$50\% in carbon-rich
stars~\cite{whitedwarf1,whitedwarf2}, but much less
(1-2\%~\cite{uranus1,Ross,uranus2,uranus3}) in Uranus and Neptune.
To give a reference point, it is useful to estimate an upper bound
to the diamond nucleation rate by considering the rate at which
diamonds would form in a hypothetical environment of pure,
metastable liquid carbon. To this end we use our numerical data on
the chemical potential of liquid carbon and diamond and our
numerical estimate of the diamond-liquid surface free energy, to
estimate the nucleation barrier of diamond as a function of
temperature and pressure. We then use CNT to estimate the rate of
diamond nucleation.

To do so, we need to extend the estimate of the nucleation rate
from the triple point pressure (around 16 GPa) up to 100 GPa, and
from the melting temperatures ( $T_{m}^{A}=$~$6600$~K and
$T_{m}^{B}=5000$~K, respectively) to 35 \% under-cooling (at which
diffusion in our sample becomes negligible on the - far from
astronomical - time-scales of our simulations).  To make such an
extrapolation, we make use of
Eqn.'s~\ref{nucleationrate},~\ref{barrier}, and ~\ref{prefactor}.
The state-point dependent quantities are the solid and liquid
number densities $\rho_{L}$ and $\rho_{S}$, the self-diffusion
coefficient $D$, the surface free energy $\gamma$, the difference
in chemical potential between the liquid and the solid $\Delta
\mu$, and the critical cluster size $n^{*}$. We estimate them in
the following way: the densities are directly measured by Monte
Carlo simulations of the solid and the liquid; the self diffusion
coefficient is extrapolated assuming an Arrhenius behaviour of the
metastable liquid (see Appendix B);  the chemical potential
difference is interpolated via Eq.~\ref{deltamu} between 30 and 85
GPa. $\Delta \mu$ then also follows  by linear interpolation.
Concerning the surface free-energy, we assume that $\gamma(P,T)$
linearly depends on $c_4$, the equilibrium concentration of 4-fold
coordinated atoms at the selected state point. This quantity is
easily measured in the Monte Carlo simulations. The nucleation
barrier height is given by Eq.~\ref{barrier}, where the
geometrical factor $c$ is the same for all cluster sizes. It is
obvious that we have to make rather drastic assumptions in order
to estimate the nucleation rate in the experimentally relevant
regime. We believe that our assumptions are reasonable, but one
should not expect the resulting numbers to provide more than a
rough indication.
\begin{figure}[h!]
\begin{center}
\includegraphics[width=0.95\columnwidth,clip]{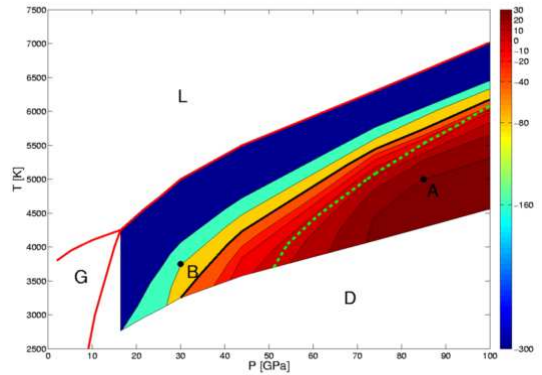}
\caption{{\small The figure shows part of the carbon phase diagram
from Ref.~\cite{Ghiringhelli05} and the iso-nucleation rate zones.
The solid red curves represent the coexistence curves from
Ref.~\cite{Ghiringhelli05}. $P_{A}=85$~GPa, $T_{A}=5000$~K and
$P_{B}=$~30~GPa, $T_{B}=3750$~K. Along the green dashed curve the
ratio of 3-fold and 4-fold coordination in the liquid is 1:1. The
numbers on the right indicate the base 10 logarithm (or the order
of magnitude) of the crystal nucleation rate from molten carbon
(in $m^{-3}s^{-1}$). The continuous black curve is the boundary of
the region above which nucleation rate becomes negligible
($<10^{-40} m^{-3}s^{-1}$).}}\label{carbonrate}
\end{center}
\end{figure}

Figure~\ref{carbonrate} shows that there is a region of some 1000K
below the freezing curve (continuous red curve) where diamond
nucleation is less than 10$^{-40}$ m$^{-3}$s$^{-1}$ (above the
continuous black line). If the rate is lower than this number, not
a single diamond could have nucleated in a Uranus-sized body
during the life of the universe. As can be seen from the figure,
our simulations for state point B are outside the regime where
observable nucleation would be expected. Note that this latter
conclusion is not based on any extrapolation.

\begin{figure}[h]
\begin{center}
\includegraphics[clip=true,width=0.8\columnwidth]{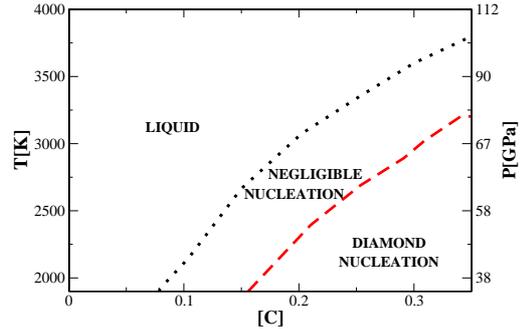}
\caption{{\small Diamond nucleation boundary as a function of
carbon concentration: in the plot, the rate is zero (no
thermodynamic driving force to nucleation) in the top region
(liquid), it is negligible ($<10^{-40} m^{-3}s^{-1}$) in the
middle region and non-negligible ($>10^{-40} m^{-3}s^{-1}$) in the
bottom-right region. We call the nucleation rate negligible if it
corresponds to less than one cluster per Uranus-sized planet over
a period of  10$^{10}$  years. The left hand y-axis represents the
temperature; the right-hand y-axis indicates the corresponding
pressure for a Uranus-like isentrope (see
Ref.~\cite{uranus1,Ross,uranus2,uranus3}).}}\label{carbonmix}
\end{center}
\end{figure}

As mentioned above, carbon stars and planets do not consist of
pure carbon. Hence, we have to consider the effect of dilution on
the crystallization process. To do so, we make a very
``conservative'' assumption, namely that nucleation takes place
from an ideal mixture of C, N,O and H~\cite{HansenBarrat}. If this
were not the case, then either demixing would occur, in which case
we are back to the previous case, or the chemical potential of
carbon in the liquid is lower than that in pure carbon, which
would imply that the thermodynamic driving force for diamond
crystallization is less than in pure liquid carbon. In
Fig.~\ref{carbonmix}, we show how dilution affects the regime
where diamond nucleation is possible. To simplify this figure, we
do not vary pressure and temperature independently but assume that
they follow the adiabatic relation that is supposed to hold along
the isentrope of Uranus~\cite{uranuscurve} and we use the
ideal-mixture expression for the chemical potential $ \beta \Delta
\mu = \beta \Delta \mu_{0} + \beta ln([C]) $, where $\beta \Delta
\mu_{0}$ is the chemical potential difference between the solid
and the liquid for he pure substance (C) and $[C]$ is the
concentration of carbon in the fluid mixture.

Not surprisingly, Fig.~\ref{carbonmix} shows that dilution of the
liquid decreases the driving force for crystallization. In fact,
no stable diamond phase is expected for carbon concentrations
below 8\%. Moreover, there is a wide range of conditions where
diamonds could form in principle, but never will in practice.
Assuming that, for a given pressure, the width of this region is
the same as in the pure $C$ case (almost certainly a serious
underestimate), we arrive at the estimate in Fig.~\ref{carbonmix}
of the region where nucleation is negligible (i.e. less than one
diamond per planet per life-of-the-universe). From this figure, we
see that quite high carbon concentrations (over 15\%) are needed
to get homogeneous diamond nucleation. Such conditions do exist in
white dwarfs, but certainly not in Uranus or Neptune.

\subsection*{Appendix A: LCBOPII}
In this appendix, we describe the main features of the latest
addition to the LCBOP family. This potential has been used in
Ref.~\cite{LCBOPII-I,LCBOPII-II,LosFasolinoNature}. However, the
simulations discussed in the present paper are based on
LCBOPI$^+$.

\begin{figure}[t!]
\centering
\includegraphics[width=0.65\columnwidth,clip]{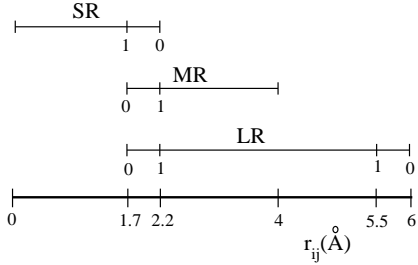}
\protect \caption{Schematic representation of the domains of the
three types of interactions. The numbers 0 and 1 below the lines
indicate the values of the switch functions as well as the
intervals where these switch functions are applied. As an example,
the SR interactions is smoothly switched off between $ r_{ij}=1.7
$ \AA~ and $ r_{ij}=2.2 $ \AA.} \label{figrev4}
\end{figure}

\begin{figure}[b!]
\centering
\includegraphics[width=0.6\columnwidth,clip]{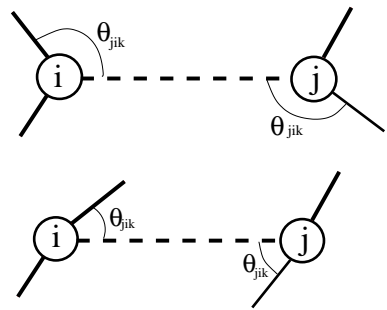}
\protect \caption{Example of a typical situations where MR
interactions are active (upper graph) and where they are switched
off by the switch function $ S_{\theta} $ (lower graph), since $
S_{\theta} $ vanishes for small bond angles $ \theta_{jik} $ and $
\theta_{ijl} $.} \label{figrev5}
\end{figure}

\begin{figure}[t!]
\begin{center}
\includegraphics[width=0.8\columnwidth,clip]{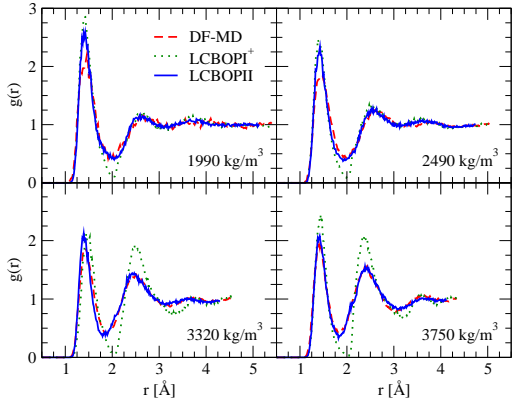}
\caption{Comparison of the radial distribution functions at 6000 K
and four selected densities between LCBOPII, LCBOPI$^+$, and the
reference data taken from our own DF-MD simulations.}
\label{rdfcomp}
\end{center}
\end{figure}

{\bf Middle range interactions}. Although LCBOPI$^+$ gave an
improved description of most liquid phase properties, like
coordination distributions as a function of density, as compared
to the bond-order potentials without LR interactions (Brenner,
CBOP\cite{Los}), the radial distribution function showed a too
marked minimum after the first shell of neighbors, as compared to
ab-initio calculations (see Fig. \ref{rdfcomp}). This deficiency
was attributed to the relatively short cut-off of 2.2 \AA~ for the
SR interactions, giving rise to a spurious barrier for bond
formation around 2.1 \AA. Therefore, for LCBOPII, the total
binding energy expression was extended with so-called MR
interactions as:
\begin{equation}
E_b =  \frac{1}{2} \sum_{i,j}^{N_{at}} \left( S^{sr}_{ij}
V^{sr}_{ij} + \left( 1 - S^{sr}_{ij} \right) V^{lr}_{ij} +
\frac{1}{\sqrt{Z^{mr}_i}} S^{mr}_{ij} V^{mr}_{ij} \right)
\label{Eb}
\end{equation}
The first two terms on the right-hand side represent the SR and LR
interactions respectively, where $ S^{sr}_{ij} $ smoothly switches
between both interactions within the interval 1.7 \AA $\leq r_{ij}
\leq$ 2.2 \AA, with $ S^{sr}(1.7) = 1 $ and $ S^{sr}(2.2) = 0 $.
The last term represents the MR interactions, where $ V^{mr}_{ij}
$ is a purely attractive potential and $ Z^{mr}_i $ is a sort of
MR coordination number defined as:
\begin{equation}
Z^{mr}_i = \frac{ \left( \sum_j S^{mr}_{ij} V^{mr}_{ij} \right)^2
} {\sum_j \left( S^{mr}_{ij} V^{mr}_{ij} \right)^2 }
\end{equation}
to account for many body effects. The switch function $
S^{mr}_{ij} = S^{mr}(r_{ij})$, going from 0 to 1 between $ r_{ij}$
= 1.7 \AA~ and $ r_{ij}$ 2.2 \AA, smoothly excludes the MR
interactions for distances smaller than 1.7 \AA. For clarity, the
ranges of the various interactions in Eq. \ref{Eb} are
schematically represented in Fig. \ref{figrev4}. The MR
interaction was fitted to ab-initio calculations of single, double
and triple bond dissociation curves. For the single bond, the tail
of the interaction vanishes beyond 4 \AA. $ V^{mr}_{ij} $ is the
product of a simple polynomial $ V^{mr}_{P,ij} $ with a smooth
cut-off at 4 \AA~ and an environment dependent switch function $
S^{mr}_{\theta} ( \{ \theta_{ijk} \} ) $, depending on the angles
between $ {\bf r}_{ij} = {\bf r}_{j} - {\bf r}_{i} $ and $ {\bf
r}_{ik} = {\bf r}_{k} - {\bf r}_{i} $ ($\forall k \neq j$), where
atom $k$ is a SR neighbour of atom $i$, i.e. $ | {\bf r}_{ik} | <
2.2 $. Thus, while the MR interactions give an extension of the
covalent interactions beyond the SR cut-off distance 2.2 \AA~ in
situations where this is appropriate, its environment dependence
relies only on the SR nearest neighbours (within 2.2 ~\AA), a
quite convenient property for the sake of efficiency. The switch $
S^{mr}_{\theta} $ acts in such a way that $ V^{mr}_{ij} $ is only
non-zero when the angles $ \theta_{ijk} $ are relatively large.
This is illustrated schematically in Fig. \ref{figrev5}. In
particular, the definition of $ S^{mr}_{\theta,ij} $ makes $
V^{mr}_{ij} $ vanish for any pair $ij$ in all bulk crystal
structures. So the addition of the MR interactions does not
require reparametrization of the SR and LR potential terms.


The reactivity of atoms depends on whether these atoms are well
surrounded by neighbours or not. Typically, an atom with a
dangling bond wants to make another bond. To include this effect,
the MR potential is made dependent on the so-called dangling
bond number $ N^{db}_i $, i.e. $ V^{mr}_{P,ij} = V^{mr}_{P,ij}
(N^{db}_i) $. For an atom with $ N^{db}_i = 1 $, the MR
interaction is stronger than for an atom with $ N^{db}_i = 0 $.

{\bf Extended coordination dependence of angular function}. For
LCBOPII the correction of the angle dependent part of the bond
order for configurations involving low coordinations and small
angles has been further extended, involving a gradual coordination
dependence of the angular term over a wide range of coordinations.

\begin{figure}[h!]
\hspace*{0.5cm}\includegraphics[width=0.9\columnwidth,clip]{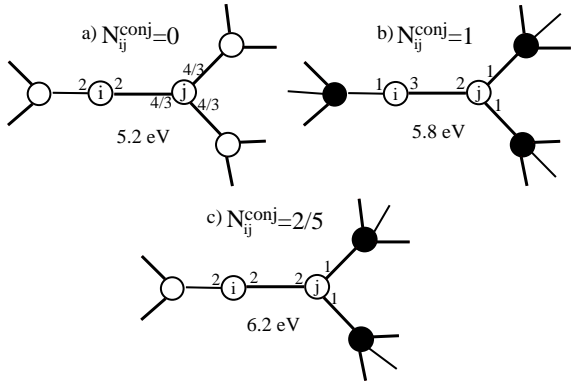}
\protect \caption{Illustration of configurations where the bond
energy is not monotonously dependent on $ N^{conj}_{ij} $, due to
the fact that bonding is less effective when the electron supply
from atom $i$, $ N^{el}_{ij} $, is not equal to that from atom
$j$, $ N^{el}_{ji} $. A correct description of each of the three
cases was achieved by adding a negative anti-bonding term to the
bond order which depends on $ \Delta N^{el}_{ij} = N^{el}_{ij} -
N^{el}_{ji} $. The resulting binding energy is given below each
configuration (see Fig. \ref{figrev3}).} \label{figrev6}
\end{figure}

{\bf Anti-bonding}. Another new feature of LCBOPII is the addition
of an anti-bonding correction to the bond order. An example of a
situation where one electron remains unpaired in a non-bonding
state is depicted in Fig. \ref{figrev6}(b). Clearly, this
situation is unfavourable as compared to the situation in Fig.
\ref{figrev6}(c). This effect cannot be captured in the
conjugation term and has therefore been included as a separate,
anti-bonding term.

{\bf Torsion}. As it has been clearly demonstrated in Ref.
\cite{Wu2002}, the torsion interaction for a bond $ij$ between an
atom $i$ and an atom $j$ is strongly dependent on conjugation,
i.e. on the coordinations of the neighbours $k (\neq j)$ of $i$
and $ l (\neq i)$ of $j$. This was already partly included in
LCBOPI$^+$. However, for LCBOPII, the conjugation dependence of
the torsion interactions was fully extended and fitted to
ab-initio calculations of the torsion barrier for all the
possible conjugation situations.

\begin{figure}[h!]
\centering
\includegraphics[width=0.8\columnwidth,clip]{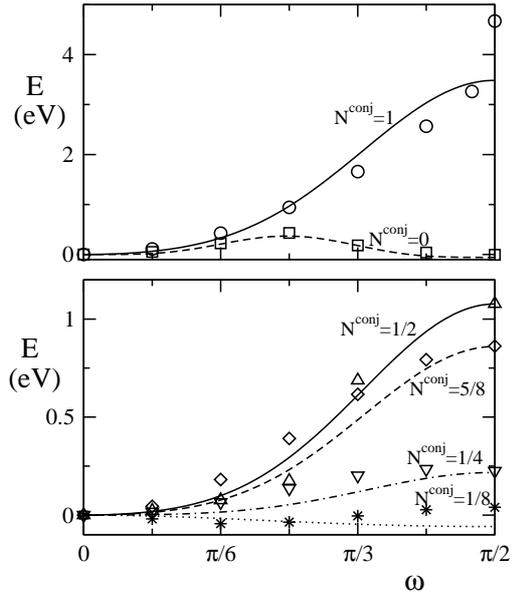}
\protect \caption{Torsional barriers according to LCBOPII and DF
calculations for the six possible values of $N^{conj}_{ij}$ for an
$ij$ bond between sp$^2$ atoms $i$ and $j$. Symbols represent the
DF results, curves the fits obtained by the LCBOPII. Top panel:
torsional barriers for the extreme values of $N^{conj}_{ij}$,
related to the conjugated ($N^{conj}$=0, squares and dashed curve)
and double bonds ($N^{conj}$=1, circles and solid curve). Bottom
panel: intermediate values of $N^{conj}$: 1/8 (stars and dotted
curve), 1/4 (down triangles and dashed-dotted curve), 1/2 (up
triangles and solid curve) and 5/8 (diamonds and dashed curve).
Note the complex behaviour of the curves for the values 1/2 and
5/8, where the barrier at $\pi/2$ is higher for $N^{conj}$=1/2
than for $N^{conj}$=5/8. The top panel alone applies also to
LCBOPI$^+$ (see Appendix A in \cite{LCBOPII-I} for details.)}
\label{figrev3}
\end{figure}

\begin{figure}[h!]
\centering
\includegraphics[width=0.8\columnwidth,clip]{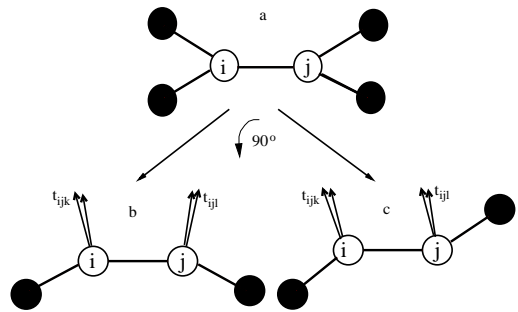}
\protect \caption{Illustration of the occurrence of spurious
torsion when using the definition of Eq. \ref{omega} (most
right-hand side) for the torsion angles.} \label{figrev7}
\end{figure}

In addition to that, LCBOPII includes a redefinition of the
torsion angle, in order to avoid the 'spurious' torsion that
occurs using the traditional definition. Traditionally, the
torsion angle $ \omega_{ijkl} $ is defined as:
\begin{equation}
cos ( \omega_{ijkl} ) = \frac{ {\bf t}_{ijk} \cdot {\bf t}_{ijl} }
{ | {\bf t}_{ijk} | | {\bf t}_{ijl} | } = \frac{ ( {\bf r}_{ij}
\times {\bf r}_{ik} ) \cdot ( {\bf r}_{ij} \times {\bf r}_{jl} ) }
{ | {\bf r}_{ij} \times {\bf r}_{ik} | | {\bf r}_{ij} \times {\bf
r}_{jl} | } \label{omega}
\end{equation}
which, assuming torsion to be non-vanishing only between sp$^2$
bonded atoms $i$ and $j$, gives rise to four torsion contributions
to the bond order. With the definition of Eq. \ref{omega}, both
situations depicted in Fig. \ref{figrev7}b and c give rise to a non-zero
torsion angle. However, in both situations there is actually no
torsional distortion but only a bending distortion, which is
already taken into account by the angular term in the bond order.
Thus, one would like to have $ \omega_{ijkl} =0 $ for the cases in
Fig. \ref{figrev7}b and c, in disagreement with the most right-hand side
expression in Eq. \ref{omega}. Another problem of expression
\ref{omega} is that it has a singularity for configurations where
$ {\bf r}_{ij} $ is parallel to $ {\bf r}_{ik} $ (or $ {\bf
r}_{il} $). For the liquid phase at high temperature such
situation are easily accessible.

For LCBOPII, the problem of 'spurious' torsion has been tackled
by a redefinition of the vectors $ {\bf t}_{ijk} $ in Eq.
\ref{omega}, reading:
\begin{eqnarray}
 {\bf t}_{ijk}  &=& \hat{\bf r}_{ij} \times ( \hat{\bf
r}_{ik_1} - \hat{\bf r}_{ik_2} ) + \\ \nonumber &+&
\frac{\sqrt{3}}{2} \left( \hat{\bf r}_{ij} \cdot ( \hat{\bf
r}_{ik_1} - \hat{\bf r}_{ik_2} ) \right) \left( \hat{\bf r}_{ij}
\times ( \hat{\bf r}_{ik_1} + \hat{\bf r}_{ik_2} ) \right)
\label{tijk}
\end{eqnarray}
and likewise for $ {\bf t}_{jil} $. Inserting these vectors into
Eq. \ref{omega}, leading to a different right-hand side,
reproduces the same $ \omega_{ijkl} $ as the traditional
definition for any torsional distortion without bending and yields
$ \omega_{ijkl} =0 $ for both situations depicted in Fig.
\ref{figrev7}, as it should be. In addition, it gives a good
interpolation for any other configuration, and it has no
singularities. Note that for the two distortions depicted in Fig.
\ref{figrev7}, the second term in Eq. \ref{tijk} vanishes, and the
vectors $ {\bf t}_{ijk} $ and $ {\bf t}_{jil} $ are parallel,
implying indeed $ \omega_{ijkl} =0 $.

\begin{figure}[h!]
\centering
\includegraphics[width=0.8\columnwidth,clip]{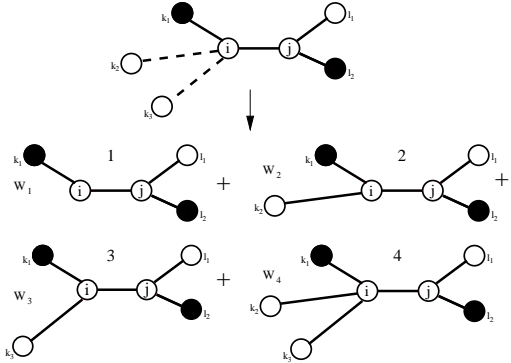}
\protect \caption{For LCBOPII, a fractional coordination situation
is treated as a weighted superpositions of integer coordination
situations, as illustrated in this picture. Dashed lines indicate
partial bonds, solid lines are for full bonds.} \label{figrev8}
\end{figure}

{\bf Interpolation for fractional coordinations}. The conjugation
term $ F^{conj}_{ij} (N_{ij},N_{ji},N^{conj}_{ij}) $ for a bond
$ij$ depends on the reduced coordinations $ N_{ij} $ and $ N_{ji}
$ of the atoms $i$ and $j$, and on the conjugation number $
N^{conj}_{ij} $. The reduced coordination $  N_{ij} $ is defined
as:
\begin{equation}
N_{ij} = \sum_{k \neq {i,j}} S_Z (r_{ik})
\end{equation}
and likewise for $ N_{ji} $, where $ S_Z (r_{ik}) $ is a switch
function for the coordination, smoothly going from 1 to 0 for $
r_{ik} $ going from 1.7 \AA~ and 2.2 \AA. $ F^{conj}_{ij} $ is
fitted to integer coordination configurations with only full
neighbours, i.e. with $ S_Z (r_{ik}) =  S_Z (r_{jl}) = 1 ~ \forall
k,l $. This poses the problem of how to determine $ F^{conj}_{ij} $
for configurations with fractional bonds, i.e. configurations
with $ S_Z (r_{ik}) < 1 $ for one or more neighbours $k$. David Brenner,
the inventor of the conjugation term, proposed to use a
3D spline~\cite{Brenner90}.
However, since the values on the integer argument nodes
are rather scattered, a spline unavoidably introduces unphysical
oscillations. For LCBOPII, we found an alternative solution to
this problem which is schematically illustrated in Fig.
\ref{figrev8}. In this approach, the conjugation term for
configurations with fractional coordination is defined as a
weighted superposition of conjugation terms for configurations
with integer configurations, the weight factors $ W_c$ for the
configuration $c$ (=1,..4) being defined in terms of the
switch functions $ S_Z (r_{ik}) $. For instance, for
the situation in Fig. \ref{figrev8}, the conjugation term
is given by:
\begin{eqnarray*}
F^{conj}_{ij} &=&
(1-S_{Z,ik_2}) (1-S_{Z,ik_3}) F^{conj}_{ij} (1,2,N^{conj}_{ij,1}) +\\
&+& S_{Z,ik_2} (1-S_{Z,ik_3}) F^{conj}_{ij} (2,2,N^{conj}_{ij,2}) +\\
&+& (1-S_{Z,ik_2}) S_{Z,ik_3} F^{conj}_{ij} (2,2,N^{conj}_{ij,3}) + \\
&+& S_{Z,ik_2} S_{Z,ik_3} F^{conj}_{ij} (3,2,N^{conj}_{ij,4})
\end{eqnarray*}
where $ N^{conj}_{ij,c} $ are the conjugation numbers for the
four configurations in Fig. \ref{figrev8}.

\textbf{Results with LCBOPII}. LCBOPII proved to be more accurate
than its predecessors in describing defects and surfaces of the
solid phases \cite{LCBOPII-I}. In the liquid phases the
improvement of LCBOPII is immediately evident when looking at
radial distribution functions at different densities (Fig.
\ref{rdfcomp}). The main discrepancy between LCBOPI$^+$ and the
reference data from DF-MD calculations
\cite{Ghiringhelli04,LCBOPII-II}) was found at the first minimum,
at around 2 \AA. LCBOPI$^+$ predicted a much deeper minimum than
DF-MD. This discrepancy is completed eliminated by LCBOPII. We
also know that the melting line of diamond predicted by LCBOPII is
about 500~K lower than for LCBOPI$^+$ at $\sim 60$~GPa
\cite{LCBOPII-II}, more consistently with ab-initio predictions of
the diamond melting line \cite{Wang2005,Correa2006}. In
\cite{LCBOPII-II} we also thoroughly analyzed the properties of
the liquid. Interestingly, by extrapolating the equations of state
of the liquid at temperatures at which our relatively small
samples actually froze, we found that a critical point for the
graphite-like into diamond-like transition is present at 1230 K.
The precise value might be inaccurate, since it is found far
outside the sampled region, still the shapes of the higher
temperature equations of state point towards the existence of such
critical isotherm. As is the case for the much speculated water
LLPT \cite{Mishima98,StanleyN} an unreachable critical point might
still be responsible of some peculiar behaviour of the system at
higher temperatures, such as the enormous change in nucleation
rate with pressure.

\subsection*{Appendix B: Self-diffusion coefficient\label{appe_diff}}

When computing the kinetic pre-factor to get the nucleation rate,
we have to consider the fact that for our model potential,
LCBOPI$^+$, only a Monte Carlo code is available. In order to
evaluate the self-diffusion coefficient needed to compute the CNT
kinetic pre-factor, we infer the scaling factor between the Monte
Carlo ``time-step'' and the MD time-step~\cite{HuitemaMCMD} by
propagating a 128 carbon atoms system via Car-Parrinello Molecular
Dynamics~\textsc{CPMD} code~\cite{Car-Parrinello} starting from a
configuration equilibrated with  LCBOPI$^+$. Note the reasonably
good agreement between the \emph{static} properties of the liquid
carbon computed with LCBOPI$^+$ and the same computed by means of
CPMD (with the BP functional)~\cite{Ghiringhelli04,LCBOPII-II}.
Data for the high pressure state point come from simulations used
in Ref.~\cite{Ghiringhelli04}, whereas data for the low pressure
state point come from a new simulation where the time rescaling is
state-point dependent, obtained with the same technical details as
reported in \cite{Ghiringhelli04}.

\begin{figure}[t!]
\begin{center}
\includegraphics[width=0.7\columnwidth,clip]{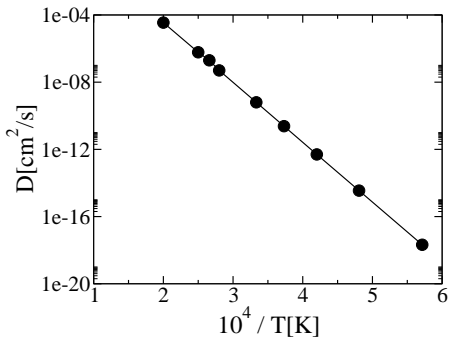}
\caption{\small Arrhenius plot for the super-cooled liquid carbon.
The activation energy is $E_{A}=7$eV from Ref.\cite{Kanter}.}
\label{diffusion}
\end{center}
\end{figure}

We use the fact that molten carbon is an Arrhenius-like liquid:
therefore, once the activation energy is known, we compute the
viscosity as a function  of temperature and by means of the
Stokes-Einstein relation  obtain the diffusion coefficient.
In the nineteen-fifties,  Kanter~\cite{Kanter} estimated the relevant
activation energy of liquid carbon to be $E_{A}=$683 $\frac{kJ}{mol}$. Subsequenty,
Fedosayev~\cite{fedosayev} reported a measurement of the molten
carbon viscosity: $\eta$ = 5 $\times 10^{11}$ poise at $T =
1860K$. We estimate the self-diffusion coefficient at the same
temperature by means of the Stokes-Einstein relation~\cite{donald}
\begin{equation}
D = \frac{k_{B} T}{ \eta a},
\end{equation}
where $a=1.54\dot{A}$ and $k_{B}$ is the Boltzmann's constant:
$D(1860 K)= 3.3 \times 10^{-17}$ cm$^{2}$/s. Since
molten carbon is an Arrhenius-like fluid~\cite{Debenedetti},
\begin{equation}
D(T) = D_{0} \exp^{-\frac{E_{A}}{k_{B}T}},
\end{equation}
we obtain  $D_{0}$: $D_{0}= 470$cm$^{2}$/s, and then
extrapolate the diffusion coefficient for different temperatures,
as shown in Fig.~\ref{diffusion} and Table~\ref{tav3}.
\begin{table}[h]
\centerline{
\begin{tabular}{||c|c||c|c||}
\hline {\bf T[K]} & {\bf D$[\dot{A}^2/s]$} & {\bf T[K]} & {\bf D$[\dot{A}^2/s]$}\\
\hline  1750      &  2.1 $\times$ $10^{-2}$  &    3000    &  6.3 $\times$ $10^{6}$  \\
\hline 2080       &  3.5 $\times$ $10^{1}$   &   3300     &  7.6 $\times$ $10^{7}$ \\
\hline 2380       &  5.0 $\times$  $10^{4}$  &   3570     &  5.0 $\times$  $10^{8}$\\
\hline 2680       &  2.4 $\times$  $10^{5}$  &    3760    &  2.0 $\times$  $10^{9}$\\
\hline
\end{tabular}}
\caption{{\small Self diffusion coefficient as a function of
temperature.}} \label{tav3}
\end{table}

We then find that at state point $A$, $D_{A} = 3.5 \times 10^{-5}$
cm$^{2}$/s, whereas at state point $B$, $D_{B} = 2 \times 10^{-7}
$  cm$^2$/s.

We also use a Car-Parrinello Molecular
Dynamics~\cite{Car-Parrinello} to calculate the self-diffusion
coefficient by means of the mean square displacement: at state
point $A$ $D=2.3 \times 10^{-5} $cm$^2$/s, which matches surprisingly  well
with the diffusion coefficient estimated by means of the Arrhenius
law, $D=3.5 \times 10^{-5} $ cm$^2$/s.

\section*{Acknowledgments}

The work of the FOM Institute is part of the research program of
FOM and is made possible by financial support from the Netherlands
Organization for Scientific Research (NWO). We gratefully
acknowledge financial support form NWO-RFBR Grant No 047.016.001
and FOM grant 01PR2070. We acknowledge support from the Stichting
Nationale Computerfaciliteiten (NCF) and the Nederlandse
Organisatie voor Wetenschappelijk Onderzoek (NWO) for the use of
supercomputer facilities. LMG wishes to thank his wife, Sara
Iacopini, for helping him in scanning and summarizing the
literature mentioned in section \ref{CPDhistory}.

{\small
\bibliographystyle{phjcp}

}
\end{document}